\documentclass[article]{aa}
\usepackage{psfig,latexsym,longtable,lscape,supertabular}

\def\arcsec{$^{\prime\prime}$\ }

\def\x{$\times$}
\def\>{$>$}
\def\<{$<$}

\def\newline{\hfil\break}
\def\mincir{\ \raise -2.truept\hbox{\rlap{\hbox{$\sim$}}\raise5.truept  
\hbox{$<$}\ }}                        
\def\magcir{\ \raise -2.truept\hbox{\rlap{\hbox{$\sim$}}\raise5.truept  
\hbox{$>$}\ }}                        

\def\ie{{\rm i.e. }}
\begin{document}

\pagenumbering{arabic}
\title{
A thorough study of the intriguing X-ray emission from the
Cartwheel ring
}

\author{Anna\,Wolter, Ginevra\,Trinchieri}
\offprints{anna@brera.mi.astro.it}

\institute{
Osservatorio Astronomico di Brera, via Brera 28, 20121
 Milano Italy}

\abstract{We present the results 
from the high resolution Chandra observation of
the \object{Cartwheel} galaxy. Many individual sources are resolved in the
image, mostly associated with the outer ring. All detected sources
have a very high X-ray luminosity ($\geq 10^{39}$ erg s$^{-1}$)
that classifies them as
Ultra Luminous X-ray sources (ULX). The brightest of them is possibly the
most luminous individual non-nuclear source observed so far, 
with L$_X \sim 10^{41}$ erg s$^{-1}$ (at D=122 Mpc). 
The spatial extent of this source is consistent 
with a point source at the Chandra resolution.  
The luminosity function of individual X-ray sources extends about
an order of magnitude higher than previously reported in other galaxies.
We discuss this in the context of the ``universal" luminosity function
for High Mass X-ray Binaries and we derive a Star Formation Rate 
higher than in other starburst galaxies
studied so far.
A diffuse component, associated with hot gas, is present. 
However, deeper
observations that we will obtain with XMM-Newton 
are needed to constrain its properties.
}

\titlerunning{The X-ray Cartwheel}

\maketitle

\section{Introduction}

The Cartwheel galaxy is a spectacular object, with the peculiar appearance
reminiscent of a wheel (hence the name), most probably the result of an 
impact with one of the companion galaxies.  
It is located in a tight, compact group (\object{SCG 0035-3357}; 
Iovino et al. 2003) of 4 members, 
very close in space ($\sim 0.3$ Mpc\footnote{We use H$\rm_0$ = 
75 km s$^{-1}$ Mpc$^{-1}$, which implies a scale of 1.252 kpc/arcsec 
at the distance of the group, D$_L$ = 122 Mpc, throughout the paper.}
and velocity $\sim 400$ km s$^{-1}$
from one another, see Taylor \& Atherton 1984). 
Whether the impact was due to G3 (at $\sim 1^{\prime}$ NE, Higdon 1996) or G2
(at $\sim 3^{\prime}$ to the North; Athanassoula, Puerari \& Bosma 1997),  
two rings are now visible as a result: the outer one has the
largest linear diameter measured in ring galaxies: 80$''$ ($\sim$100 kpc) 
along the major axis; the inner one, close to the core, is elliptical in
shape with obvious dust lanes crossing it
(Struck et al. 1996).

Many detailed observations of the Cartwheel are available, 
ranging from radio line (Higdon 1996) and continuum (Higdon 1996), 
to near- (Marcum et al. 1992) and far-infrared (Appleton \& Struck-Marcell
1987), optical (Theys \& Spiegel 1976, Fosbury \& Hawarden 1977) and 
H$\alpha$ images (Higdon 1995) and line spectroscopy (Fosbury \& 
Hawarden 1977). All have confirmed
the presence of a recent starburst in the outer ring, without a
corresponding activity in the inner ring, nucleus or spokes, believed
to be relatively devoid of gas.   Most 
of the activity is confined in fact in the S-SW portion of the ring,
where massive and luminous HII regions characterized by large
H$\alpha$ luminosities and equivalent widths are found (Higdon 1995).
Both dynamical considerations and stellar evolution models suggest 
an age of 2-4 \x 10$^8$ yr for the star burst. The estimated supernova
rate, as high as 1 SN/yr (\ie almost two orders of magnitude higher than in 
normal galaxies), coupled with the evidence of a very low metallicity
environment (as measured from O, N and Ne)
also supports the view that star formation in the ring is a
recent phenomenon and that the gas currently forming stars 
was nearly primordial at the time
of the impact (Fosbury \& Hawarden 1977; Higdon 1995; Marcum et al. 1992).  

We have imaged the Cartwheel for the first time in the X-ray band,
using the HRI on board ROSAT (Wolter, Trinchieri, Iovino 1999), and 
were able to attribute most of the emission to the outer ring, 
stronger in the Southern quadrant and very clumpy in nature.
We therefore asked and obtained Chandra data with the ACIS-S 
in imaging mode, to better study the spatial distribution of the 
emission. 
We present here the Chandra data, and a discussion of the detection
of both a diffuse component and a number of Ultra--Luminous X-ray sources
well in excess of expectations.

The paper is organized as follows: in Sect. 2 we present the Chandra data;
in Sect. 3 the results of the analysis for the individual sources and
for the extended component; in Sect. 4 we discuss the results and
present the X-ray Luminosity Function of the Cartwheel sources.
Sect. 5 summarizes our findings.

\section{Chandra data}

\begin{figure*}
\resizebox{18cm}{!}{
\psfig{file=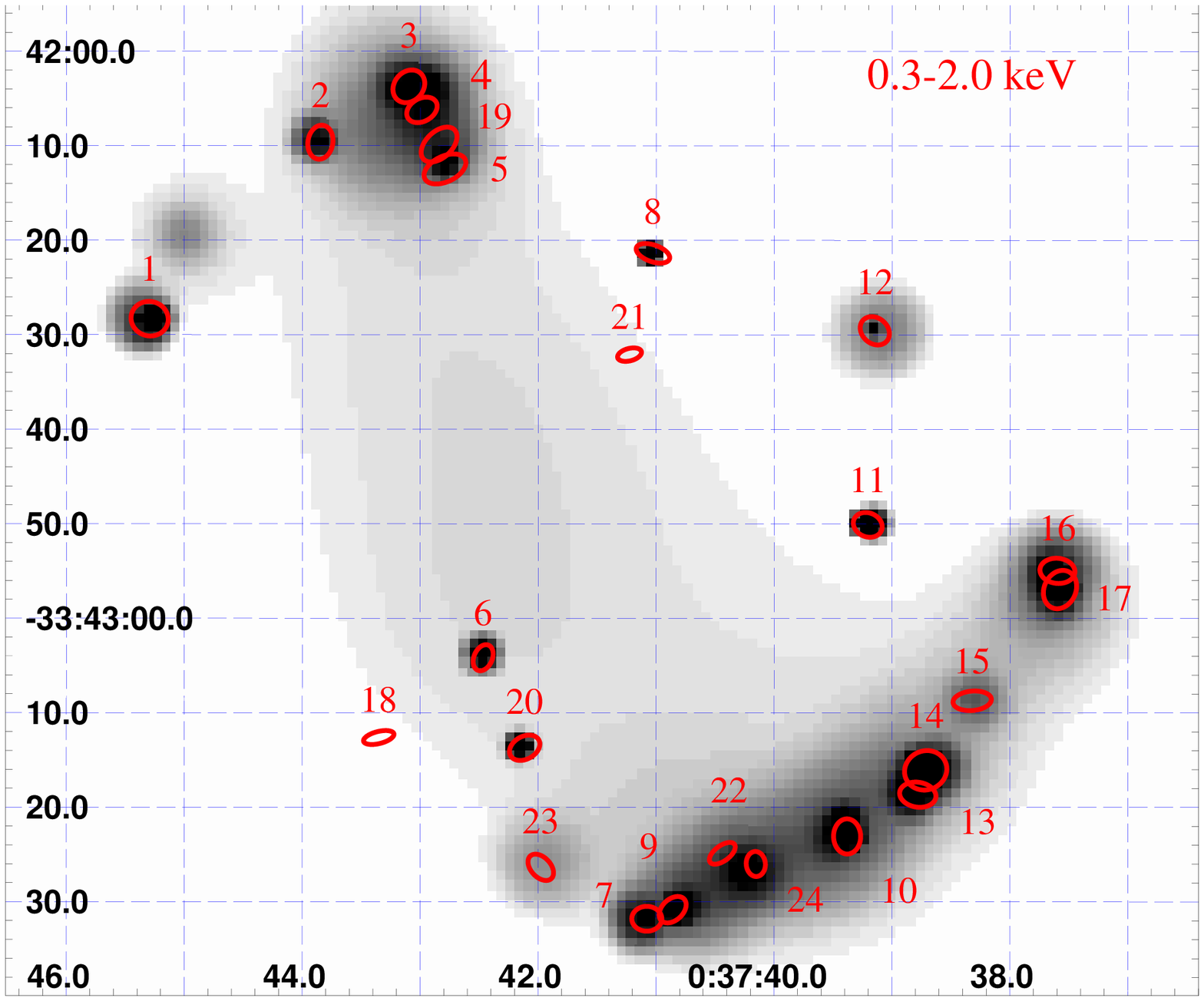,width=9cm}
\psfig{file=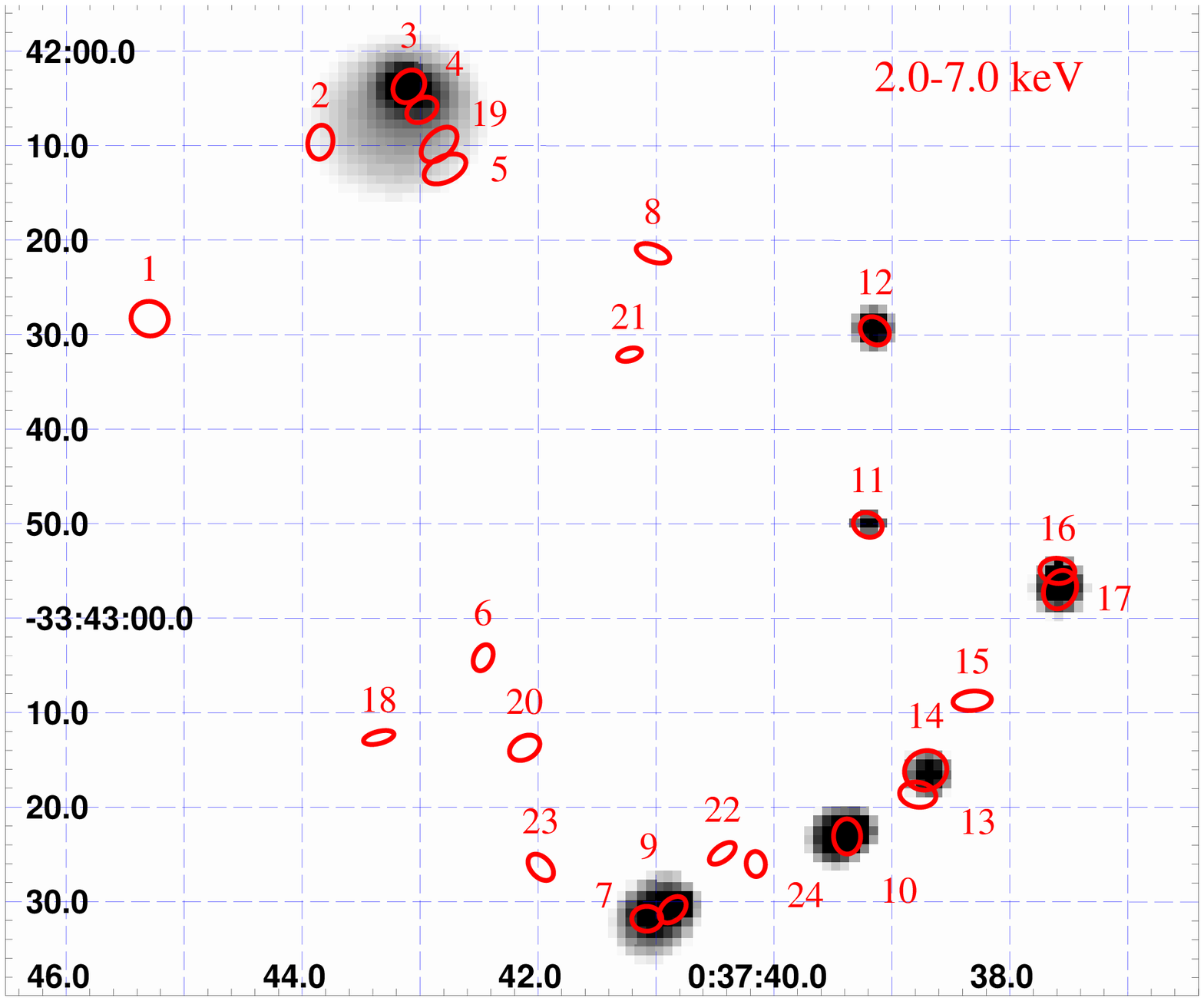,width=9cm}
}
\caption{Chandra image of the Cartwheel in the ({\it Left})
soft (0.3-2 keV) and ({\it Right}) hard (2.0-7.0 keV) energy bands.
An adaptive smoothing algorithm has been applied to the data (see text).
Numbers identify sources as given in Table~\ref{tab1}, with regions 
used to derive the source counts (from wavdetect).}
\label{chhard}
\end{figure*}

The Cartwheel was observed by Chandra on 26-27th May 2001, 
with ACIS-S in imaging mode, operated in the standard full-frame mode,
with an integration time of 3.2 s. The back-illuminated ACIS-S S3 chip 
was chosen for its soft response to detect the low temperature component.
Telemetry was in faint mode.
For a description of the Chandra mission see Weisskopf et al. (2000).

The data were reduced with the standard Chandra pipeline with
the CIAO software (version 2.3) and the most recent available calibration
products, as described in \verb+http://asc.harvard.edu/+.
The corrections applied are those appropriate for the ACIS-S instrument.

No evidence of in-orbit high background was found in the data, so the 
net exposure time is 76.1 ksec. 

Fig.~\ref{chhard} shows the adaptively smoothed images of 
the Cartwheel
in two energy bands (0.3-2.0 keV and 2.0-7.0 keV respectively; 
the energy limits are chosen as a good compromise that maximizes
the signal and minimizes 
the particle background contribution and calibration uncertainties).
Both images show a very clumpy emission, mostly confined to the outer ring
(see also Wolter et al. 1999, Wolter \& Trinchieri 2003 and Gao et al. 2003).
Individual sources that appear point-like at the Chandra resolution
($\leq 1.5$ kpc at the Cartwheel distance, see Sect.~\ref{extent}) account
for most of it, in particular in the harder energy band (right panel). 
A fraction of the counts ($\sim 20\%$ of the total) is found in a
more extended component, that appears in the softer energy band. 
An additional component, of very low 
surface brightness, appears to connect the southern portion of the
ring to the two nearby companions G1 \& G2 in the soft image 
(left panel of Fig.~\ref{chhard}).
No excess emission is detected at the position of the optical nucleus.

\begin{figure}
\psfig{file=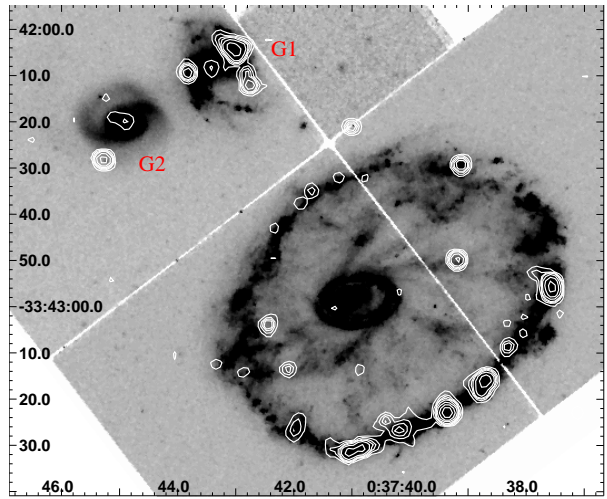,width=8truecm,clip=}
\caption{The X-ray contours in the 0.3-7 keV Chandra image
over-plotted on the optical image in the F450W filter from the HST archives.
X-ray contours are: 7.9$\times 10^{-06}$, 1.6$\times 10^{-05}$, 
2.4$\times 10^{-05}$, 3.2$\times 10^{-05}$, 5.3$\times 10^{-05}$, 
1.1$\times 10^{-04}$, 2.1$\times 10^{-04}$, 4.2$\times 10^{-04}$ 
cnts/sec/arcsec$^2$}
\label{opt}
\end{figure}

The X-ray contours from the smoothed 0.3-7 keV image are shown in 
Fig.~\ref{opt} superposed on to the 
HST image~\footnote{The  F450W filter image was obtained  
from the HST WFPC2
association archive. The image is based on observations made 
with the NASA/ESA Hubble Space Telescope, obtained from the data 
archive at the Space Telescope Science Institute. STScI is operated by the Association of Universities for Research in Astronomy, Inc. under NASA contract NAS 5-26555.}
and show the correspondence of the X-ray emission with the outer ring.

We will explore all the evidence more quantitatively in the following sections.

\section{Analysis and Results }
\subsection{Individual sources}

A {\em wavdetect} detection algorithm applied to the 0.3-7.0 keV image
provides 72 sources in the whole field. We have applied the algorithm
using scales ranging from 1$^{\prime\prime}$ to  1$^{\prime}$ and
a significance threshold of $10^{-6}$ that corresponds to 0.25
false sources in the image considered at any given scale.
All detected sources are listed and discussed in the Appendix and in 
Table~\ref{wavsrc}.  

In Table~\ref{tab1} we list
the positions and count rates (in the 0.3-7.0 keV band) of the 25
sources located in the Cartwheel region, most of which should be
associated with it (see also Fig.~\ref{chhard} and Sect. 2).

By assuming the spectrum found for the sum of point sources 
(see Sect.~\ref{secspe}), we compute the unabsorbed fluxes in the 0.5-2 keV
and in the 2-10 keV bands. 
Assuming the distance of 122 Mpc for the Cartwheel we list in Table~\ref{tab1}
the logarithm of the k-corrected
luminosity L$_X$ in the 2-10 keV band.

The brightest point source in the outer ring has a luminosity of at least
L$_X \sim 2-4 \times 10^{40}$ erg s$^{-1}$ in the 0.5-2. keV
band, and of L$_X \sim 5-9 \times 10^{40}$ erg s$^{-1}$ in the 2-10 keV band
\footnote{The luminosity depends on the spectral model assumed, see Table~\ref{fit}.}, one
of the brightest individual sources ever found in 
galaxies.
Even though at the Cartwheel distance the Chandra resolution defines
a region of a few kpc that could contain more than one source, 
the high X-ray luminosity suggests either a single extremely
bright source, or a very dense collection of several, high L$_X$ sources, 
which is probably even more peculiar. A possible source variability suggested
for this source (see Sect. 4.1) further suggests that this is indeed
a single high L$_X$ source 
(or equivalently that a single source dominates the emission from this
region).

\begin{table*}

\caption{List of sources that are detected in the area covered by the
Cartwheel system and displayed in 
Fig.~\ref{chhard}. Unabsorbed fluxes are computed in the 0.5-2 keV 
and 2-10 keV bands assuming the best fit model of the sum of point
sources (see text for details). 
Unabsorbed, k-corrected luminosities are in the 2-10 keV
band. 
The sources are also reported in the
Appendix in RA order for completeness; however, in the Appendix the fluxes 
are all computed under a different spectral hypothesis (see text). 
These sources are identified by their number in the last column of 
Table~\ref{wavsrc} for easy reference.}
\begin{tabular}{| r r r r r r r r|}
\hline
N & Position (J2000)    & Net Counts &   Count rate & F(0.5 -2 keV) & F(2-10 keV) &log L$_X$ & Notes\\
  &                    &                    & $\times 10^{3}$ sec & erg cm$^{-2}$ s$^{-1}$    & erg cm$^{-2}$ s$^{-1}$   &  2-10 keV  &\# as in Gao\\
\hline
 1 &00:37:45.294 -33:42:28.31&  73.21$\pm$ 8.77 &  0.961$\pm$0.115 & 3.59$\times 10^{-15}$ & 2.94$\times 10^{-15}$ &39.72 &a;\#30\\
2 &00:37:43.847 -33:42:09.63&  39.87$\pm$ 6.56 & 0.524$\pm$0.086 & 1.96$\times 10^{-15}$ & 1.60$\times 10^{-15}$ &39.45 &a;b;\#29\\
3 &00:37:43.117 -33:42:03.96&  72.32$\pm$ 8.72 & 0.950$\pm$0.115 & 3.55$\times 10^{-15}$ & 2.90$\times 10^{-15}$ &39.71 &a;b;\#24\\
4 &00:37:43.006 -33:42:05.96&  48.56$\pm$ 7.14 & 0.638$\pm$0.094 & 2.38$\times 10^{-15}$ & 1.95$\times 10^{-15}$ &39.54 &a;b;\#25\\
5 &00:37:42.791 -33:42:12.46&  40.58$\pm$ 6.63 & 0.533$\pm$0.087 & 1.99$\times 10^{-15}$ & 1.63$\times 10^{-15}$ &39.46 &a;b;\#22\\
 6 &00:37:42.466 -33:43:04.18&  19.97$\pm$ 4.69 &  0.262$\pm$0.062 & 9.79$\times 10^{-16}$ & 8.02$\times 10^{-16}$ &39.15 &\#21\\
 7 &00:37:41.078 -33:43:31.80&  70.64$\pm$ 8.60 &  0.928$\pm$0.113 & 3.46$\times 10^{-15}$ & 2.84$\times 10^{-15}$ &39.70 &\#17\\
 8 &00:37:41.028 -33:42:21.38&  21.23$\pm$ 4.80 &  0.279$\pm$0.063 & 1.04$\times 10^{-15}$ & 8.52$\times 10^{-16}$ &39.18 &a;\#16\\
 9 &00:37:40.859 -33:43:30.84&  66.20$\pm$ 8.31 &  0.869$\pm$0.109 & 3.25$\times 10^{-15}$ & 2.66$\times 10^{-15}$ &39.67 &\#15\\
10 &00:37:39.380 -33:43:23.08& 383.77$\pm$19.77 & 5.040$\pm$0.260 & 1.88$\times 10^{-14}$ & 1.54$\times 10^{-14}$ &40.44 &\#11\\
11 &00:37:39.206 -33:42:50.11&  62.59$\pm$ 8.06 & 0.822$\pm$0.106 & 3.07$\times 10^{-15}$ & 2.51$\times 10^{-15}$ &[39.65] &\#10\\
12 &00:37:39.147 -33:42:29.57&  43.75$\pm$ 6.78 & 0.575$\pm$0.089 & 2.15$\times 10^{-15}$ & 1.76$\times 10^{-15}$ &39.49 &\#9\\
13 &00:37:38.781 -33:43:18.66&  38.02$\pm$ 6.40 & 0.499$\pm$0.084 & 1.87$\times 10^{-15}$ & 1.53$\times 10^{-15}$ &39.43 &\#7\\
14 &00:37:38.714 -33:43:16.08& 105.92$\pm$10.54 & 1.391$\pm$0.138 & 5.19$\times 10^{-15}$ & 4.25$\times 10^{-15}$ &39.88 &\#6\\
15 &00:37:38.320 -33:43:08.72&  16.06$\pm$ 4.24 & 0.211$\pm$0.056 & 7.88$\times 10^{-16}$ & 6.45$\times 10^{-16}$ &39.06 &\#5\\
16 &00:37:37.598 -33:42:54.96&  94.51$\pm$ 9.90 & 1.241$\pm$0.130 & 4.64$\times 10^{-15}$ & 3.79$\times 10^{-15}$ &39.83 &\#3\\
17 &00:37:37.576 -33:42:56.94& 116.10$\pm$10.95 & 1.525$\pm$0.144 & 5.70$\times 10^{-15}$ & 4.66$\times 10^{-15}$ &39.92 &\#2\\
18 &00:37:43.351 -33:43:12.63&   6.82$\pm$ 2.83 & 0.090$\pm$0.037 & 3.35$\times 10^{-16}$ & 2.74$\times 10^{-16}$ &38.69 &a\\
19 &00:37:42.841 -33:42:09.63& 18.57$\pm$ 4.58 & 0.244$\pm$0.060 & 9.11$\times 10^{-16}$ & 7.46$\times 10^{-16}$ &39.12 &a;b;\#23\\
20 &00:37:42.114 -33:43:13.71&  13.20$\pm$ 3.87 & 0.173$\pm$0.051 & 6.47$\times 10^{-16}$ & 5.30$\times 10^{-16}$ &38.97 &\#20\\
21 &00:37:41.226 -33:42:32.10&   6.97$\pm$ 2.83 & 0.092$\pm$0.037 & 3.41$\times 10^{-16}$ & 2.80$\times 10^{-16}$ &38.70 & \\
22 &00:37:40.438 -33:43:24.87&  14.28$\pm$ 4.00 & 0.188$\pm$0.053 & 7.00$\times 10^{-16}$ & 5.73$\times 10^{-16}$ &[39.01] &\#13\\
23 &00:37:41.988 -33:43:26.36&   8.54$\pm$ 3.16 & 0.112$\pm$0.042 & 4.19$\times 10^{-16}$ & 3.43$\times 10^{-16}$ &38.78 &\#19\\
24 &00:37:40.154 -33:43:26.00&   5.16$\pm$ 2.45 & 0.068$\pm$0.032 & 2.53$\times 10^{-16}$ & 2.07$\times 10^{-16}$ &38.57 &\#12\\
\hline

\end{tabular}
N.B.: Sources \#4, \#8, \#18 and \#27 in Gao et al. (2003) 
are not detected by us. 
Our sources N.18 and N.21 are not detected by Gao et al.

Luminosities in [ ] indicate sources that could be modulated by higher absorption because they appear to have a soft/hard count ratio different from the others.

a): not used to compute the Luminosity Function.

b): associated with G1; G2 is below our detection threshold.

\label{tab1}
\end{table*}

No source is detected at the position of the optical nucleus of the
Cartwheel, indicating that it is fainter than our detection limit
of a few $\times 10^{38}$ erg s$^{-1}$.   An active nucleus is thus either 
not present or it is so heavily absorbed that even at $\sim$7 keV no emission
emerges.

A number of individual sources coincide with G1, for a total
of 220 net counts, that correspond to a flux f$_{(0.5-2.0 keV)}$  = 1.1 
$\times 10^{-14}$ erg cm$^{-2}$ s$^{-1}$ and
f$_{(2.0-10.0 keV)}$ 
= 8.8$\times 10^{-15}$ erg cm$^{-2}$ s$^{-1}$; for a luminosity 
L$_X$=2.5$\times 10^{40}$ erg s$^{-1}$ (0.5-10 keV band)
(assuming the same spectrum of the sum of point sources, see Fig.~\ref{sp1}
and Sect. 3.1.2).
The disturbed optical morphology of G1 and the large 
number of discrete high L$_X$ sources detected in the galaxy area
are both consistent with an interaction scenario (G1 with G2, as suggested
by Higdon 1996).

A source that might be positionally coincident with the nucleus
is also detected at the location of 
G3 at 3.9 $\sigma$ with a total of 9.6$\pm$3.3 counts (see Appendix A). 
The corresponding flux obtained with the same spectrum as for G1
is f$_{(0.5-2.0 keV)}$=4.6$\times 10^{-16}$ erg cm$^{-2}$ s$^{-1}$; 
f$_{(2.0-10.0 keV)}$=3.9$\times 10^{-16}$ erg cm$^{-2}$ s$^{-1}$; for
a total luminosity L$_X$=1.5$\times 10^{39}$ erg s$^{-1}$  (0.5-10 keV band).

While we were preparing this paper, a list of sources in the Cartwheel
region was presented by Gao et al. (2003). We compare our list
with that presented by them. Positions are generally consistent,
however there is a small systematic shift of \<$\Delta$RA\> =
$-0.25^{\prime\prime}$ in RA and \<$\Delta$Dec\> =
$-0.60^{\prime\prime}$ in declination between our positions and those
of Gao et al.  A few sources are only present in either list (see 
notes to Table~\ref{tab1}):  it is likely that 
differences in the detection process and thresholds assumed cause the slightly
different source lists;  the effect is however limited to faint or 
confused sources.   
The count rates of the common sources are generally higher in the Gao et al.
list. This might be the result of different bands used (the detection band
is not indicated in the Gao et al. paper) and/or background 
subtraction considered and/or a different detection algorithm
(under the IDL software).

\begin{figure*}
\resizebox{18cm}{!}{
\psfig{file=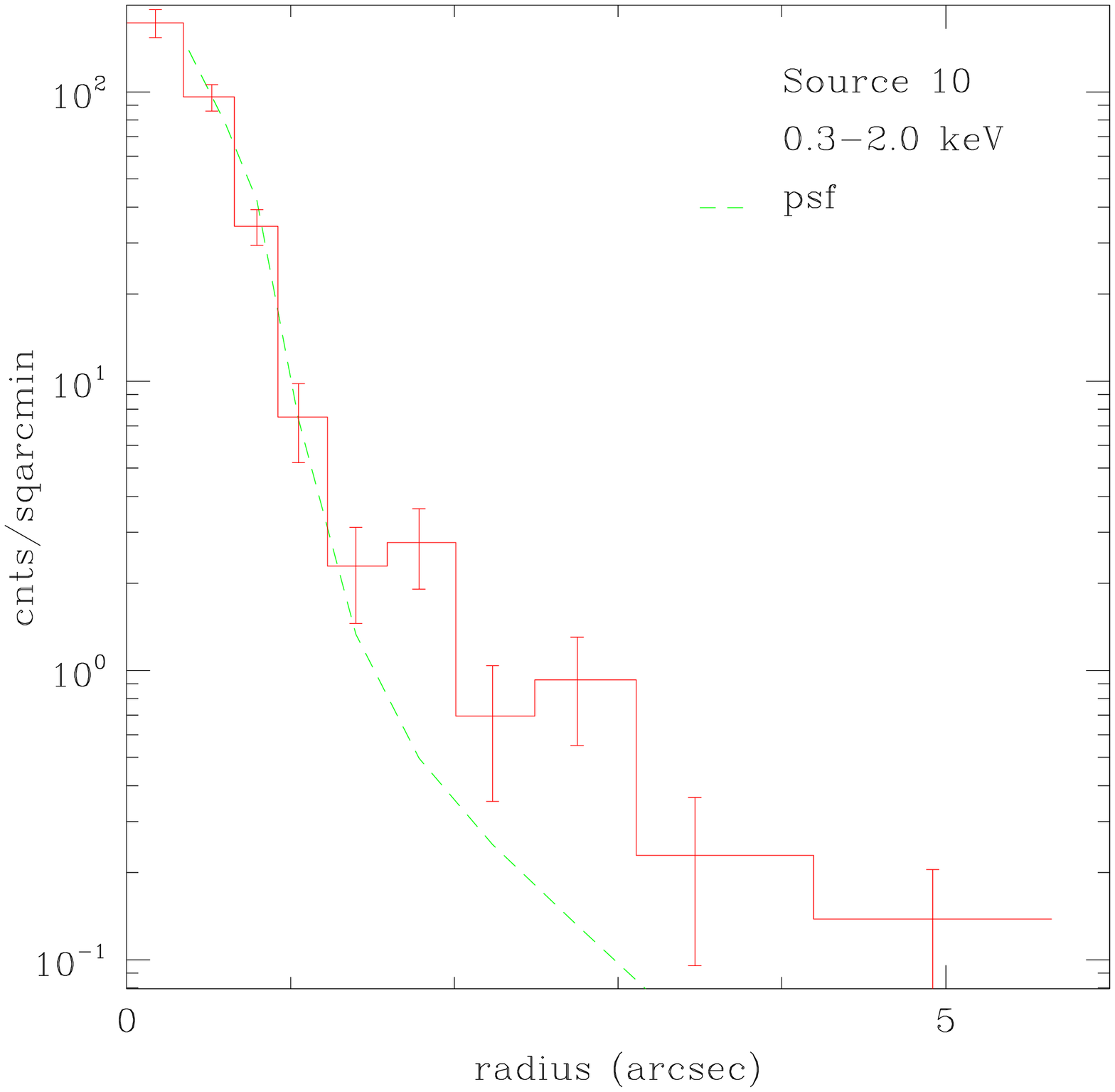,width=18truecm,height=14truecm}
\psfig{file=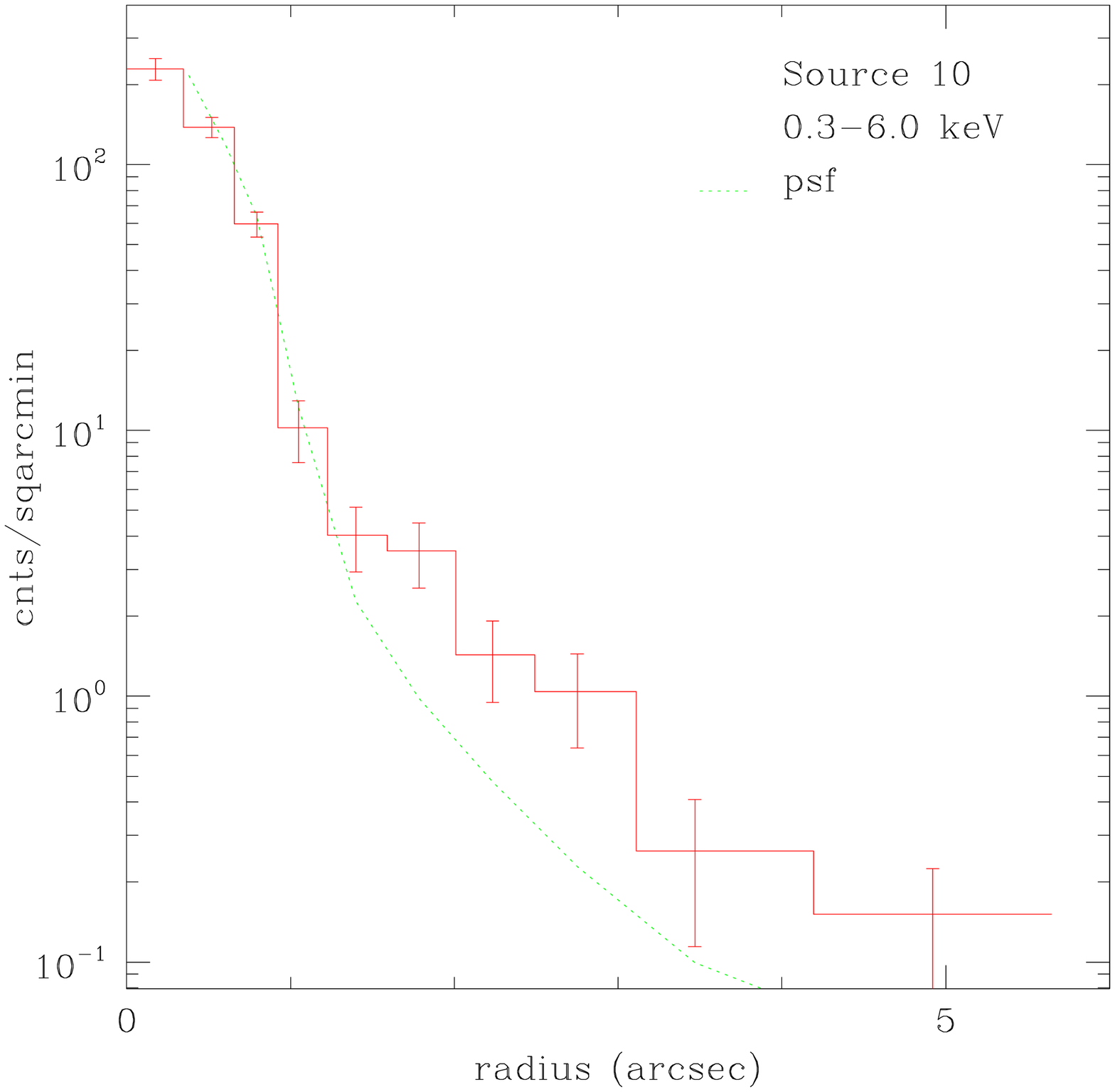,width=18truecm,height=14truecm}
}
\resizebox{18cm}{!}{
\psfig{file=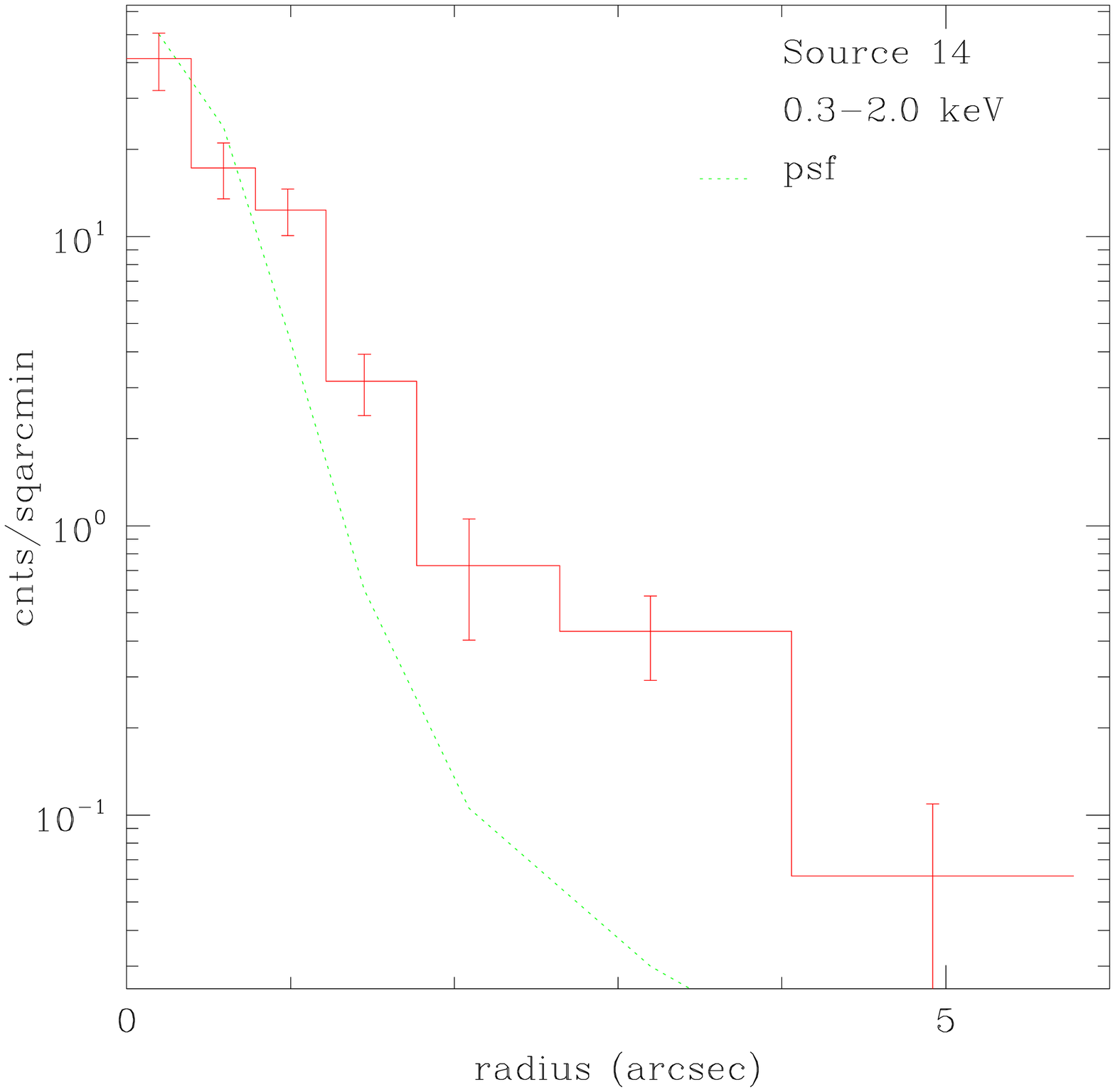,width=18truecm,height=14truecm}
\psfig{file=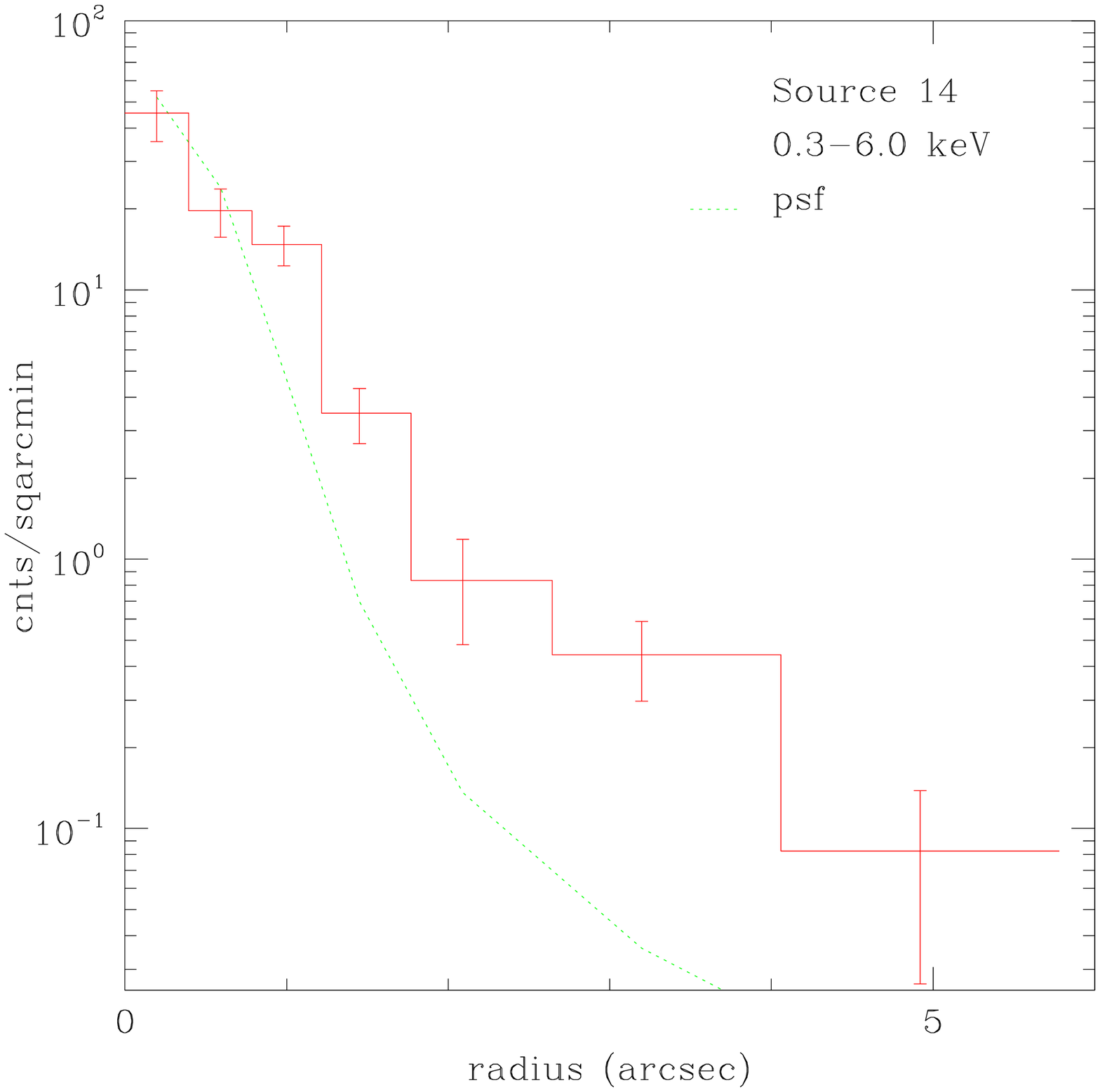,width=18truecm,height=14truecm}
}
\caption{ Surface brightness profile of source N.10
in the soft range (0.3-2. keV) {\bf Top left} and in the total
range (0.3-6. keV) {\bf Top right} compared with the PSF from {\em chart} 
in the same energy ranges.
Surface brightness profile of source N.14
in the energy range 0.3-2. keV ({\bf Bottom left})
and 0.3-6. keV ({\bf Bottom right}) compared with the PSF from {\em chart}
in the same energy ranges. Source N.13 has been masked out before computing
the profiles.
}
\label{profiles}
\end{figure*}

\subsubsection{Extent}
\label{extent}

In order to investigate the nature of the sources we have studied the
radial profile of the brightest ones (\ie with sufficient counts to perform 
a significant comparison with the Point Spread Function, PSF, obtained 
by ray tracing the photons distributed with the same spectral shape as the 
source under test with the CIAO task {\em chart}).
The profiles of two of these source are plotted in Fig.~\ref{profiles}
in two different energy bands, after subtracting the detected nearby sources,
together with the PSF binned in the same energy range.

The sources are unresolved at the resolution of the instrument, \ie they 
have a core of $\mincir 1^{\prime\prime}$. An extended source of lower surface
brightness is present and it is stronger at soft energies for 
$r \geq 1\farcs5$, most evident for source N.10.
We infer that 
this is due to the underlying ring of star formation in the Cartwheel.
The thickness of this component, $\sim 8-10^{\prime\prime}$, is
comparable to that in the optical band.

\begin{figure*}
\psfig{file=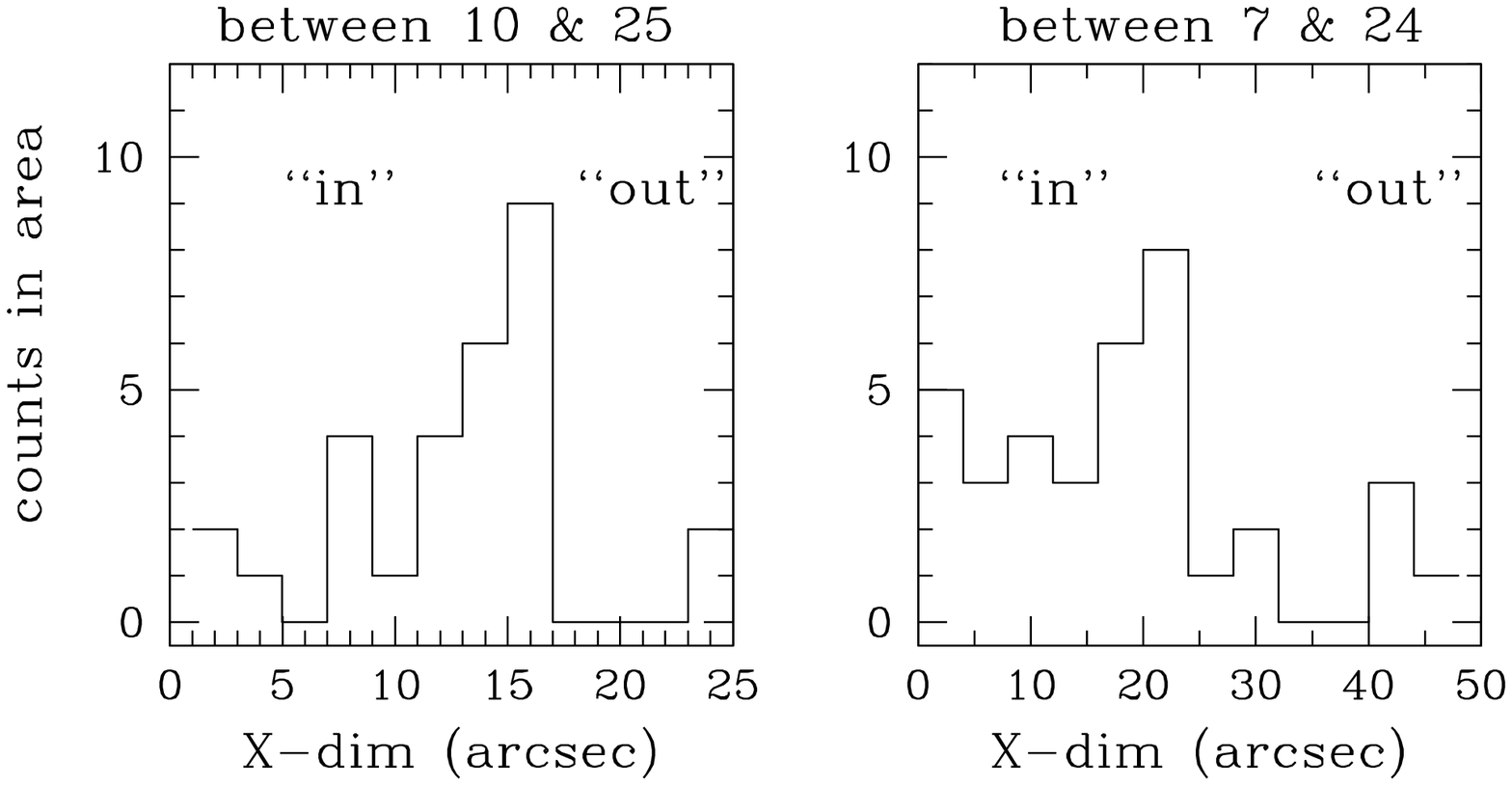,width=16truecm,clip=}
\caption{Cuts in regions orthogonal to the star-forming ring. See text for comments.
The regions used are slits of 5.6$^{\prime\prime}$ and 5.9$^{\prime\prime}$
(left and right panels respectively) and length given by the x-axis. 
{\it Left}: In the gap between the brightest source N.10, and source N.25, 
\ie approximately in the South direction; 
{\it Right}: In the gap between sources N.7 and N.24, \ie approximately 
oriented NW-SE. 
}
\label{proj}
\end{figure*}

We have also performed a number of projections,  orthogonal to the ring,
in a few spots of the southern ring. We plot two of these in
Fig.~\ref{proj}. The location is given relative to the sources
identified in Fig.~\ref{chhard}.
The X-axis is approximately along the radial distance
from the center of the ring, from the inside towards
the outside of the Cartwheel.
The peak in the distribution corresponds to the ring, which appears as
an annular plateau over which point sources stand out. 
These plots also suggest that there is a higher level of
emission ``inside'' the ring (\ie low ``X-dim'' values) 
than ``outside'' (\ie high ``X-dim'' values) and that
a rather sharp drop confines the outer limb of the ring.
With the limited number of counts detected ``inside" the ring, detailed studies
of this component are not possible and will have to wait for more sensitive
observations.

\subsubsection{Spectrum}
\label{secspe}

The spectrum of the brightest source in the Cartwheel ring has 
already been 
described by Gao et al. (2003). However, with our analysis we find
slightly different results; unfortunately not enough details
are given in the Gao et al. paper for us to ascertain if the
differences are based on different hypotheses or different treatment
of the data. Therefore we present here in detail the spectral analysis
of the brightest source (the only one for which a spectral analysis can
be attempted, with $\sim$380 net counts) and for the sum of the other 
individual sources detected around the ring (for which we collect in total
more than 600 net counts). 
The background has been computed in large circles
devoid of bright sources at about the same off-axis position as the Cartwheel.

\begin{table}

\caption{Spectral fit results for source N.10}.
\begin{tabular}{| l r r|}
\hline
&\multicolumn{1}{c}{Power Law }&\multicolumn{1}{c|}{Multicolor Disk} \\
\hline\hline
N$_{\rm H}$ $\times 10^{21}$ cm$^{-2}$&3.6(2.6-5.9) &2.4(0.6-3.4) \\
$\Gamma$/ kT &1.6 (1.3--2.0) &1.3 (1.0-2.1)\\
$\chi^2$ (dof)& 9.97 (9) & 9.73 (9) \\
&&\\
F$_x$ (0.5-2 keV) obs. & 1.2$\times 10^{-14}$ & 1.2$\times 10^{-14}$ \\
F$_x$ (0.5-2 keV) unabs.  & 2.5$\times 10^{-14}$ & 1.8$\times 10^{-14}$\\
F$_x$ (2-10 keV) obs. & 5.0$\times 10^{-14}$ & 2.9$\times 10^{-14}$ \\
F$_x$ (2-10 keV) unabs.  & 5.2$\times 10^{-14}$& 5.3$\times 10^{-14}$\\
&&\\
L$_x$ (0.5-2 keV)& 4.3$\times 10^{40}$ & 2.1$\times 10^{40}$ \\
L$_x$ (2-10 keV) & 9.1$\times 10^{40}$ & 5.3$\times 10^{40}$ \\
L$_x$ (0.05-100 keV) & 4.5$\times 10^{41}$& 9.$\times 10^{40}$ \\
\hline\hline
\end{tabular}
\label{fit}
\end{table}

\begin{figure}
\psfig{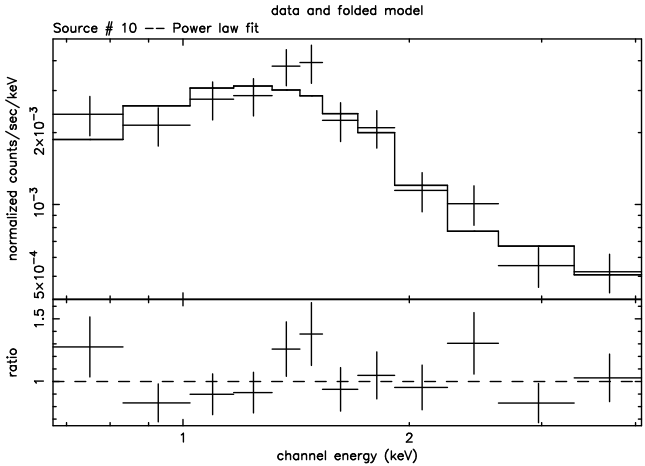}
\caption{ACIS-S spectrum of source N.10.
The solid line corresponds to a power law model 
with $\Gamma=1.6$  and low energy absorption 
N$_H= 3.6 \times 10^{21}$ cm$^{-2}$ ({\em wabs+pow}). 
The lower panel shows the ratio between data and model.}
\label{s10pl}
\end{figure}

For the brightest source (N.10) we bin the data to have at least 30 total
counts in each bin, and we apply a simple power-law model with low
energy absorption (see Table~\ref{fit}).  
The fitted N$_{\rm H}$ value is consistent with other absorption measures
in the Carthweel: the best fit is higher than the line-of-sigth
Galactic value but consistent with the intrinsic absorption
measured e.g. in the optical band:
the value of A$_V$=2 measured (Fosbury \& Hawarden 1977) 
corresponds to N$_{\rm H} = 3.8 \times 10^{21}$ cm$^{-2}$
(using A$_V$ = N$_{\rm H} 
\times 5.3 \times 10^{-22}$, for R$_V$ = 3.1, see eg. Bohlin, Savage \& 
Drake, 1978) well within the range of the fitted N$_{\rm H}$ values
in Table~\ref{fit}.
The power law slope would indicate a High Mass X-ray Binary (HMXB), as also 
proposed by Wolter
et al. (1999) based on luminosity arguments.

Although a more complex model is not required by the data, we also used 
the  absorbed
Multicolor Disk model ({\em diskbb} model in XSPEC; Mitsuda et al. 1984)
used for other ULX (see eg. Bauer et al. 2003, Zezas et al. 2002),  
which gives an equally good fit 
(see Table~\ref{fit}). 
The formal errors in the temperature/N$_{\rm H}$ parameters (90\% confidence region for one
interesting parameter) are given in Table~\ref{fit}.

Unfortunately, given the statistics,  a
thermal plasma model, like a Raymond-Smith or a Mekal with fixed
solar abundance also give formally
acceptable fits ($\chi^{2}_{\nu}$ of the order of 1.1), with a temperature 
($\magcir 2.5$ keV)  not strongly constrained.
Therefore we feel that more detailed spectral models will not convey
meaningful additional information for this source. 
However, we notice the same small excess at 
$\sim$ 1.5 keV that was pointed out by Gao et al. (2003).
This line could correspond to a feature from either Mg, or Al, but if these
are due to a low temperature plasma, there should be more prominent lines
around 1 keV.
Lines produced by low ionization states of Mg, Si, and S in an optically thin gas 
(see eg. Iwasawa et al. 2003) could also be an explanation:
e.g. in the spectrum of source \#11 in the Antennae 
a MgXIII line is fitted at
1.50 keV (Zezas et al. 2002). 
While adding a Gaussian component would clearly reduce
the minimum  $\chi^{2}$, 
we feel that this is not required, either 
statistically (the $\chi^{2}_{\nu}$ is about 1 even without
the line) or from the distribution of the residuals.
 Moreover it is also clear from the Gao et al. results that their 
 $\chi^{2}_{\nu}$ are small ($<1$ always), reflecting the very low 
 significance of each bin. 

Fluxes and luminosities are given in Table~\ref{fit} for different bands,
including a luminosity in the 0.05 - 100 keV range, 
often assumed to be a measure
of the bolometric luminosity.  All luminosities are k-corrected.
As the values in Table~\ref{fit} show, the ``bolometric'' luminosity
strongly depends on the model and on the N$_{H}$ value assumed,
given the large extrapolation to both low and high energies, 
therefore caution should be used when comparing ``bolometric''
luminosities derived 
from data extracted in significantly smaller energy ranges and
extrapolated using different models.

To derive an average spectrum for the individual sources in the
Cartwheel we also accumulated the counts from 
$\sim 1^{\prime\prime}$ circles around the positions associated 
to the other individual sources detected.
Source N.10 is not included since it is by far the brightest and its
inclusion would heavily bias the results.  We also
do not include sources N.11 and N.22 that appear to have a 
lower soft/hard count ratio than the average source, 
suggesting either an intrinsically different 
spectrum or a larger absorbing column. 
Since their location is also not coincident with the star-forming ring,
they could be background sources seen through
the absorbing material in the Cartwheel. Optical observations (imaging
and spectral) are needed to confirm the identification.

The spectral data are binned so that each bin has a 
significance $\geq 2 \sigma$ after background subtraction.
The resulting spectrum is shown in
Fig.~6. The data are fitted by a power law with best fit values 
N$_{\rm H}$ = 1.9 [1.5-2.3] $\times 10^{21}$ cm$^{-2}$ 
and $\Gamma$ = 2.2 [2.04-2.34] (reduced $\chi^2$ = 1.12 for 50 dof). 
Even in this case, the derived low energy absorption is higher than galactic,
but consistent with the reddening observed in the HII regions.
The contour plot of the uncertainties in spectral index and
low energy absorption is shown in Fig.~\ref{conf}. 
The photon index we obtain is steeper than what is observed in other bright 
binaries in nearby galaxies when fitted with a simple power law model, 
and in particular steeper than the spectrum of source N.10, but
a few examples of steep sources are present e.g. in the Antennae 
(Zezas et al. 2002) or 
in other galaxies (e.g. Humphrey et al. 2003).
Since it is likely that this slope results from a combination of different 
slopes, and perhaps even of different intrinsic absorption, depending on the 
location of the source with respect to the star forming region, this 
result should be taken with caution and might not be indicative of
a population of steeper spectra sources.

The measured unabsorbed flux is 3.0/2.6 $\times 10^{-14}$ erg cm$^{-2}$ s$^{-1}$ 
(in the 0.5-2.0/2.0-10 keV band),
which corresponds to a total luminosity of 
5.3$\times 10^{40}$/4.6$\times 10^{40}$ (0.5-2.0/2.0-10.0 keV) erg s$^{-1}$. 
The sum of the luminosity of the detected point sources is then roughly
twice that of N.10 in the soft band and comparable to that of N.10 in the hard band.

\begin{figure}
\psfig{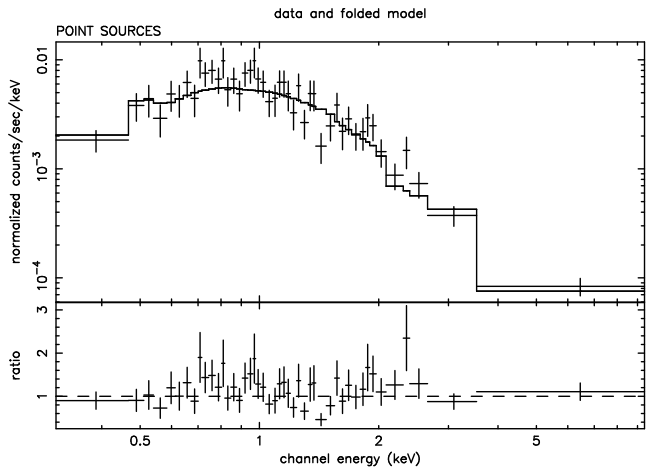}
\caption{ACIS-S spectrum of the combined individual
sources. The solid line corresponds with a power law fit with
$\Gamma=2.2$ and  N$_{\rm H}$ = 1.9 $\times 10^{21}$
cm$^{-2}$. Source N.10 is not included since it would strongly
bias the statistics. Sources N.11 and N.22 are also
not considered since they have a very different spectral distribution
possibly indicating that they are either background sources or intrinsically
absorbed ones.}
\label{sp1}
\end{figure}

\begin{figure}
\psfig{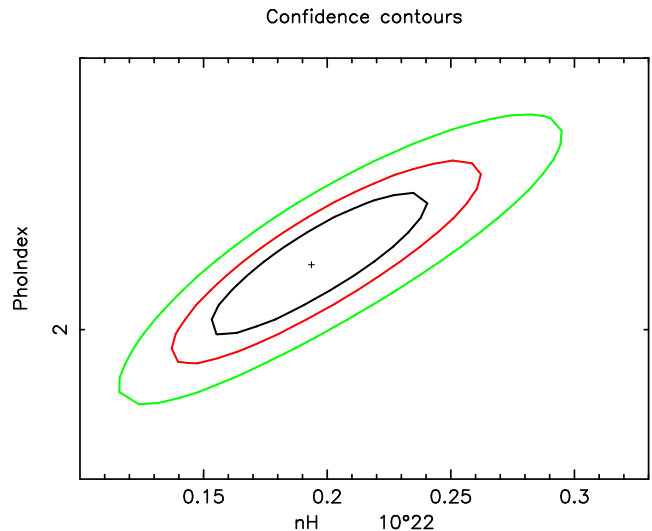}
\caption{Confidence contours of the two parameters $\Gamma$ and N$_{\rm H}$
for the combined point-source spectrum. Curves are at the 68\%, 90\% and 99\%
confidence level for the two parameters.}
\label{conf}
\end{figure}

\subsection{Diffuse component}
\label{diff}
As mentioned above, we have also detected a more extended component
that coincides
with the ring. There is a slight indication of diffuse emission also
in the center of the galaxy, but with no direct relation to the other
optical structures like the inner ring or the spokes.
The smoothed image in the soft band (see Fig.~\ref{chhard}) suggests also a
low surface brightness emission extending towards galaxies G1 and  G2,
however, the statistical significance 
is low. Support for the presence of a diffuse component comes from the 
plots in Fig.~\ref{proj}, 
where the average emission inside the ring is higher than outside it. 
Fig.~\ref{profiles} also shows a shoulder at the ring position more 
pronounced
in the softer than in the harder band. We conclude therefore that there
might be a hot gaseous component that permeates the Cartwheel and might
be enhanced, or more heated, in the ring.

\begin{figure}
\psfig{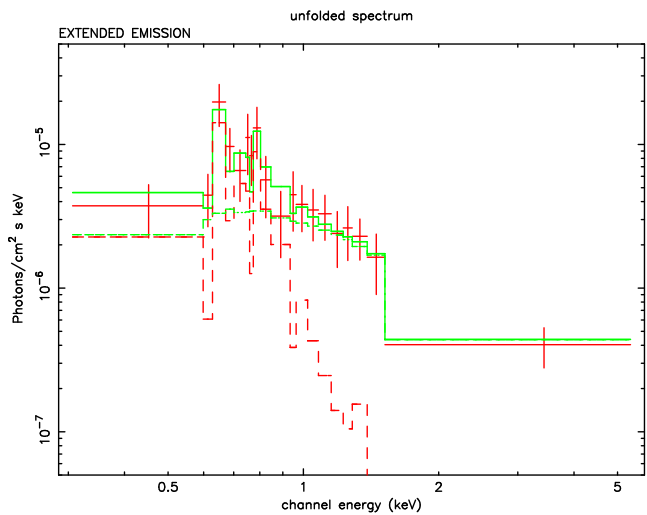}
\caption{
ACIS-S unfolded spectrum of the diffuse component:
even if the statistics are low, the spectrum cannot be fitted by a single
component. We show here the fit with a hot plasma component (dashed line) plus
power law (dashed gray line) plus the total fit (solid gray line). See text 
for details. 
}
\label{spunf}
\end{figure}

To investigate the spectral properties of the ``extended'' emission
we collect counts from a region that includes virtually all of the 
Cartwheel extent, but excludes all detected point sources analyzed
previously. 
The spectrum  has been binned to have a significance $\geq 2 \sigma$ after
background subtraction in each bin.
In spite of the limited statistics,
we found that the spectrum requires at
least two components.
A single component fit (e.g. power law, plasma model with fixed abundance)
is formally acceptable since the errorbars are large, but it has badly 
distributed residuals.
We tried different combinations of two components, which are all
equally acceptable, given the statistics.
A good representation (Fig.~\ref{spunf}) is
given by  a combination of a power law 
($\Gamma$=2.3, N$_{\rm H}$=2.3$\times 10^{21}$ cm$^{-2}$)
and a plasma model (Raymond-Smith component with  kT=0.2 keV and
an abundance fixed at 0.5$\times$ solar to reflect the low metallicity 
of the gas measured in optical-IR.).
The power law component, which could represent fainter unresolved individual
binaries, has a slope consistent with that found for the combined point 
sources (see Sect. 4.1.3). The unabsorbed flux of this component is 
F$_x$ = 1.1$\times 10^{-14}$/8.5$\times 10^{-15}$ erg cm$^{-2}$ s$^{-1}$ (0.5-2.0/2.0-10.0 keV),
about 25/10\% of the resolved point source flux in the soft and hard band 
respectively.
The flux of the diffuse hot gas component is F$_x$ = 1.6 $\times 10^{-14}$
erg cm$^{-2}$ s$^{-1}$ (0.5-2.0 keV band).
It contributes mostly at 0.6-0.9 keV as expected from the temperature
found.  The total luminosity of the gas is of the order of 
$\sim 3 \times 10^{40}$ erg s$^{-1}$, \ie only about a factor 4 less than the 
soft gaseous component in the most X-ray luminous starburst galaxy known, 
NGC 3256 (Moran et al. 1999).

\section{Discussion}

\subsection{Comparison with literature results}

In Fig. 9 
we show the HRI contours overlaid on the adaptively
smoothed Chandra image in the soft band. 
The only striking difference in the image is in source labeled ``A'' in
the figure, 
which is not present in the new Chandra image. However,
the source is probably not related to the Cartwheel itself, so it might be 
any kind of variable source from an AGN to a transient Galactic source.
The total soft luminosity
seen by Chandra (adding up all the different components, from point sources
to diffuse gas) 
of L$_X$ (0.2-2.0) $\sim 2\times 10^{41}$ erg s$^{-1}$ is consistent with that of 
the HRI of  $2.5 \times 10^{41}$ cgs, once the same  H$_0$ is assumed,
and taking into account the different 
spectral parameters used to derive the luminosities.

A strict comparison of the contribution from each individual source is 
not easy, given the very different size of the PSF of the two instruments. 
However, 
by comparing the main knots of emission in the HRI with the corresponding
regions in Chandra, 
we notice a variation in the flux level in at least one region.
If we make the hypothesis that the region around G1 has not varied,
then the region around sources N.10-13-14 shows a factor of $\sim 3$ increase.
This is consistent with variability observed in other ULX,
if one source only is responsible for the flux increase.
Since N.10 is the dominant source we can probably attribute the
variation to this one source; otherwise, as already suggested in 
Sect. 3.1, we should assume that there is
a large compact cluster of sources that vary together in a region
of $\sim 1.5 - 2$ kpc. For comparison, in a similar region in 
the Antennae, closer and therfore more resolved, 
there are up to  7 sources brighter than 10$^{38}$ erg s$^{-1}$, of which 
only 2 are above  10$^{39}$ erg s$^{-1}$ (see 
Fig. 1 and Table 1 of Zezas et al. 2002), which would contribute 
only about 1/10 of the observed luminosity of source N.10.

The flux variation implies that an origin of the X-ray emission from SNR is 
probably less likely than the association with an accreting compact
object.

\begin{figure}
\psfig{file=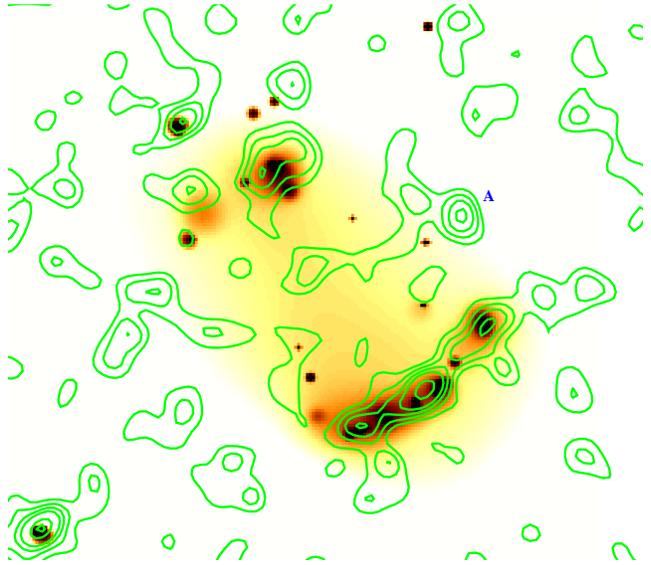,width=8.5truecm}
\caption{
The ROSAT-HRI contours from the adaptively smoothed image
are over-plotted onto the Chandra soft adaptively smoothed image
in gray scale.
}
\label{hri}
\end{figure}

Even if N.10 is by far the brightest source in the ring, it is clear
from Fig.~\ref{hri} that the HRI was also detecting the entire
ring emission. 
We estimate in fact that  in the 0.2-2.0 keV band,  
the brightest
source N.10 contributes only $\sim$ 1/4 of the total luminosity.
Gao et al. (2003) instead propose that most 
of the emission detected by the ROSAT HRI is from this source only;
however they use a much broader band than the 0.2-2.0 keV that 
matches the ROSAT HRI energy band.

\begin{figure}
\psfig{file=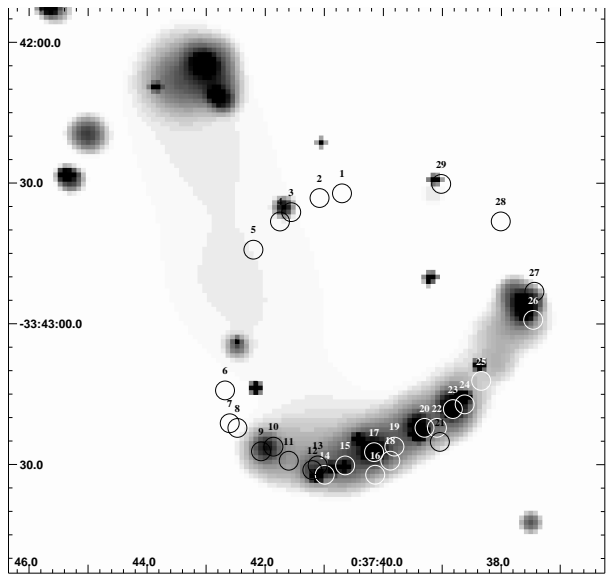,width=8truecm,clip=}
\caption{X-ray smoothed map in the (0.3-7.0) keV band. 
Over-plotted are the position of HII regions 
numbered as in Higdon(1995).}
\label{hii}
\end{figure}

Giant HII regions and complex structures, typically coincident with peaks
of H$\alpha$ emission, have been observed in actively star-forming objects 
like the interacting system ``The Antennae'' (NGC 4038/9; e.g. Fabbiano, 
Zezas \& Murray, 2001) as extremely bright X-ray sources, with intrinsic
luminosities reaching several $\times 10^{40}$ erg s$^{-1}$. To check this association
in the Cartwheel, we have plotted the positions of the HII knots as 
measured by Higdon (1995) on the X-ray image in Fig.~\ref{hii}. 
The positions of the circles that mark the HII regions have been
shifted 
by about 1$^{\prime\prime}$ in RA and 0.5$^{\prime\prime}$ 
in Dec for better agreement with the locus of 
the X-ray peaks (well within the positional uncertainty of both
the X-ray and the HII reference frame). A general trend in the location
of the X-ray and HII emission is evident. However, there is no one-to-one 
correspondence with any of the X-ray sources. If anything, the X-ray emission
seems to be at the edge of the HII knots.
The same kind of general positional agreement is evident between the
X-ray emission and the Mid-IR peaks
(Charmandaris et al. 1999); the so called ``hot spot'', especially at 
15$\mu$m, dominates the output in this energy range. It coincides with
two large HII complexes, but not with the brightest X-ray source.
The resolution of the ISO data is however 6-7$^{\prime\prime}$, so more
than one X-ray source (e.g. at least N.22 and N.24) might be
associated with the ISO emission. 
The best interpretation of this similarity is that the region is active 
in general, but time scales and regions of emission
are not directly linked. 

\subsection{Individual sources}

The excessive number of very high X-ray luminosity individual sources makes 
it interesting and complicated to understand their nature.

The limiting luminosity for point sources 
is of the order of $5 \times 10^{38}$ erg s$^{-1}$, 
which is already above the Eddington limit for a neutron star binary 
($L_X \sim 3\times 10^{38}$ erg s$^{-1}$; see e.g. King et al. 2001).
Most of the detected sources are also above the limits of canonical
Ultra-Luminous X-ray sources (ULX), \ie $L_X \geq 10^{39}$ erg s$^{-1}$. 
The main uncertainty in the $L_X$ calculation is the correct association of a
few of the individual sources to the Cartwheel itself (\ie 
they might be foreground or background sources). We assume that all sources
within the optical ring belong to the Cartwheel. However
there are also a number of sources not positionally 
coincident with the  ring: N.6, N.8, N.11, N.20, N.22
that we consider below one by one.

N.8 is located between the Cartwheel and the G1 galaxy, so it is
probably unrelated to either source, unless the  encounter 
affected even
this area. We consider this unlikely and treat it as a
background source.

From comparison of the soft and hard count rates  sources 
N.11 and N.22  appear to have a different spectrum from the rest. 
If this results from higher absorption, with  
the limited statistics available,  we estimate
an absorber of $\leq 10^{22}$ cm$^{-2}$, consistent with a
galaxy like the Cartwheel itself.
However it is not possible to determine whether the sources
are embedded in the absorber, and therefore belong to the Cartwheel,
or are behind it, and therefore background sources. Given the
location of N.22 in the ring we consider this to belong to the
Cartwheel.  The association of N.11 is less certain, however by analogy
with N.22 we also consider it as part of the galaxy and include both in 
the estimate of the Luminosity Function in the next section.

N.6 and N. 20 are not exactly on the peak of star formation, but close to 
the inner side of the ring. We consider them ``leftovers'' from past 
star formation, and therefore related to the Cartwheel.

We discuss below the properties of the ULX in the Cartwheel on a 
statistical basis, by deriving their Luminosity Function.

\subsection{Luminosity Function of ULX in the Cartwheel ring}

We have computed the luminosity function (LF) of all 16 isolated sources
detected along the outer ring, assuming the distance of the Cartwheel
for computing the luminosities (sources that are not used are indicated with
a note in Table~\ref{tab1}.)

\begin{figure}
\psfig{file=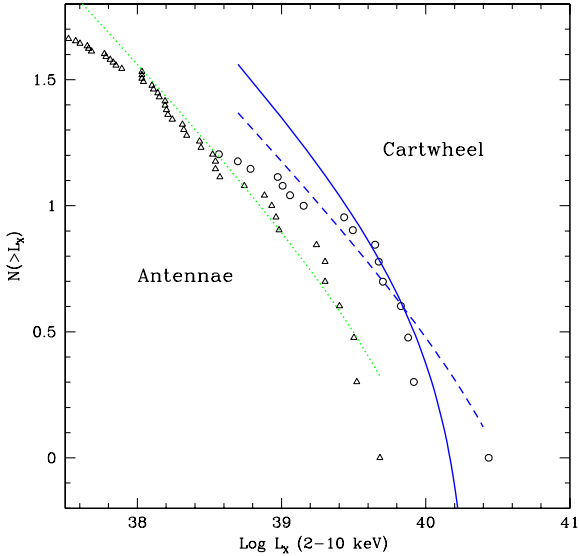,width=8truecm}
\caption{Luminosity Function of all bright isolated sources in the ring.
The solid line represents the Grimm et al. (2003) luminosity function with SFR=20
$M_{\odot}$ yr$^{-1}$ and cut-off luminosity of L$_X = 2\times 10^{40}$ erg s$^{-1}$. 
The dashed line has SFR=12$M_{\odot}$ yr$^{-1}$ and
cut-off luminosity of L$_X = 1\times 10^{41}$ erg s$^{-1}$. 
The Antennae LF from Zezas 
et al. (2002) is also plotted for comparison, with the
original fit from Grimm et al. (2003).  We have assumed a constant factor
of 0.5 to convert luminosities from the original 0.1-10 keV band to the 
2-10 keV band considered here. }
\label{lumfun}
\end{figure}

Grimm, Gilfanov \& Sunyaev (2003) propose that
a ``universal'' LF for HMXB can be constructed by normalizing to
the Star Formation Rate (SFR)
in the formula $$N(>L) = 5.4 \  SFR \  (L_{38}^{-0.61} -210^{-0.61})$$
(see their Eq. (7)).
The resulting differential LF has a slope of $\alpha$=1.61 and a cut-off
luminosity of L$_X$ (2-10 keV) = 2.1$\times 10^{40}$ erg s$^{-1}$.
We plot the histogram of the LF in Fig.~\ref{lumfun} and compare it 
with the expectation from the 
Grimm et al. (2003) formula with a value of SFR = 20 M$_{\odot}$ yr$^{-1}$,  
and with the Antennae LF (from Zezas et al. 2002) with the original
fit of the Grimm et al. (2003) formula.

We notice: 1) The slope 1.61 reproduces reasonably well the distribution, 
above L$_x \sim 10^{39}$ erg s$^{-1}$;
2) the low luminosity flattening  might be due to incompleteness
(lower flux sources are harder and harder to detect
above a diffuse emission plateau, see e.g. discussion in Kim \& Fabbiano,
2003);
3) the cut-off luminosity should be higher than in the Grimm et al. formula:
at least source N.10 is above the assumed cut-off;
4) with a cut-off at L$_X \sim 1 \times 10^{41}$ erg s$^{-1}$, if nothing else
should change in the functional form of the above equation, the nominal 
SFR derived is $\sim 12 M_{\odot}$ yr$^{-1}$;
5) discarding source N.10 a resonable fit is obtained with the Grimm et
al formula with an SFR = $\sim 20 M_{\odot}$ yr$^{-1}$; the discrepancy
at low luminosities however is larger.

Low Mass X-ray Binaries (LMXB) are unlikely to contribute significantly at
these luminosities, since their average LF (Gilfanov, 2004) is very steep
above a few $10^{37}$ erg s$^{-1}$ and has a cut-off 
at L$_x \sim 2\times 10^{39}$ erg s$^{-1}$, so that
their contribution in the luminosity range probed by this observation would be
marginal relative to HMXB.

By following the different approach suggested by Grimm et al. (2003) of
relating the SFR to the total X-ray luminosity (their Eq. 21) we derived
a slightly higher value of $\sim 25 M_{\odot}$ yr$^{-1}$.
These results  compare to the values derived earlier, based on the efficiency
of star forming galaxies of producing X-rays (Wolter et al. 1999) of 
$ \dot{M} \sim 20 -  50 \ M_{\odot}$ yr$^{-1}$.

The higher SFR might explain the difference in luminosity of the sources found
in the Cartwheel with respect e.g. to those in the Antennae, that have
an SFR of 7.1 M$_{\odot}$ yr$^{-1}$ (Zezas et al. 2002; Fig.~\ref{lumfun}). 
The question of a higher cut-off L$_X$  in this system remains open. 

\section{Summary and Conclusion}

We have presented  results from a Chandra observation of the
Cartwheel. A number of isolated and very luminous
sources is present, closely related to the
region of high star formation detected at other wavelengths, accounting for 
at least 75-80\% of the total luminosity of L$_{X}$ = 2.2$\times 10^{41}$
erg s$^{-1}$ in the 0.5-10 keV band. 

A more extended gaseous component 
coincident with the ring is also detected, and also a more tentative 
diffuse component that might permeate the entire system.
The extended component has a low temperature (kT $\sim$ 0.2 keV) consistent
with an origin related to starburst superwinds as in NGC 3256 (Moran et
al. 1999, Lira et al. 2002) or in NGC 253 (Pietsch et al. 2001, Strickland et al. 2000) and a luminosity L$_X \sim 3 \times 10^{40}$ erg s$^{-1}$ in the
0.5-10 keV band.

Individual sources are consistent with being pointlike, although even with
the superb Chandra resolution we only probe the kpc scale at the 
Cartwheel distance.
However several considerations prompt us to suggest that we are really
detecting individual very bright sources, among the brightest 
ULX seen in external galaxies. 
The most luminous, N.10, has a 
(0.5-2/2-10 keV) L$_X > 2./7.
\times 10^{40}$ erg s$^{-1}$, to be compared with e.g. the brightest ULX 
in M82 (L$_X \sim 9\times 10^{40}$ erg s$^{-1}$ in 0.5-10 keV at its brightest; 
Matsumoto et al. 2001 and Kaaret et al. 2001), 
in NGC 4559
(L$_X = 2.\times 10^{40}$ erg s$^{-1}$ in 0.3-10 keV band; Soria et al. 2004)
or in NGC2276 (L$_X = 3.\times 10^{40}$ erg s$^{-1}$ in the 0.5-2 keV band;
Davis \& Mushotzky, 2004).
All these luminosities are computed assuming isotropic emission;
the true luminosity might be lower if the X-rays are beamed. 
The crude spectral analysis of the brightest source
indicates that models typical of accreting binaries (like 
an absorbed power law or a multicolor disk model) could describe the
data. Further, a possible flux variation strongly suggests
that the observed emission is due to a single object.

The total luminosity of the individual sources reaches at least 
L$_X = 1.8 \times 10^{41}$ erg s$^{-1}$ in the 0.5-10.0 keV band.
The derived Luminosity Function is consistent with a population of
HMXB in an actively star forming object and the system requires an 
SFR of at least 12-20  M$_{\odot}$ yr$^{-1}$, in agreement with 
previous estimates for the Cartwheel.
We find a good match with the assumption of the ``universal'' 
Luminosity Function for HMXB of Grimm et al. (2003); however we suggest a
cut-off luminosity $\sim 5 \times$ higher.

The nature of ULX is still not clear and the wide range of their
characteristics points
to the possibility that they are a heterogeneous class, as indicated
by close scrutiny of nearby objects (e.g. Roberts et al. 2004). 
We are inclined to exclude SNR because of possible long term 
variability. Also, 
the spectral properties are consistent with HMXB. Since the luminosity is so
high, a black hole accreting object is more likely than a neutron star. 
However, no fit with MD gives a low enough temperature (the best fit is 
kT=1.3 keV) to indicate a very high mass compact object.

Our forthcoming XMM-Newton observation will allow us to confirm
the presence of the diffuse component with higher statistical
significance and to better define the nature of the individual sources
through spectral analysis.

\begin{acknowledgements}
We thank the referee for comments that greatly improved the presentation 
of this work.
We have received partial financial support from the Italian
Space Agency (ASI).  This research has made use of the NASA/IPAC
Extragalactic Database (NED) which is operated by the Jet Propulsion
Laboratory, California Institute of Technology, under contract with
the National Aeronautics and Space Administration.
This research has made use of SAOImage DS9, developed by Smithsonian
Astrophysical Observatory.

\end{acknowledgements}

\appendix
\section{} 
Table~\ref{wavsrc} lists all sources detected in the ACIS-S CCD7 field
detected by the wavelet algorithm (see Sect. 3.1).  
The table lists the CXO name, 
the X-ray position, the net counts and their errors from the wavelet analysis,
the significance of detection and the flux in the 0.5-10 keV band
for all the sources detected in the entire field of view of the
CCD (S3). 
Fluxes are computed for all sources by assuming Galactic 
N$_{\rm H}$ = 2.$\times 10^{20}$ cm$^{-2}$
and a power law with index $\Gamma$=1.7. 
Table~\ref{wavsrc} also includes the sources already presented
in Table~\ref{tab1} since the spectral assumptions in the two tables
are different; 
notes in the last column 
indicate the number the sources have in Table~\ref{tab1}
and other names from NED.

The wavelet algorithm detects 72 sources in the S3 CCD area 
(8.4\arcsec $\times$ 8.4\arcsec), 47 of which are not positionally
related to the Cartwheel group. 
The density of the sources corresponds therefore to $\sim 2.3 \times 10^{3}$ 
sources/sq.deg at the faintest detected flux of f$_X = 3.0 \times 10^{-16}$
erg cm$^{-2}$ s$^{-1}$ in the 0.5-10 keV band. This is entirely consistent with the density
measured in the deep Chandra surveys (e.g. Rosati et al. 2002).

None of these field sources is identified in the literature although several
optical/radio associations are possible  (e.g. PKS or NVSS sources,
faint optical counterpart visible on the POSS, or even with magnitude from
APM, etc.)
We discuss here briefly the X-ray properties of the brighter ones, for which
we have explored the spectral properties 
(i.e. those sources with more than 500 net counts).

\begin{figure}
\psfig{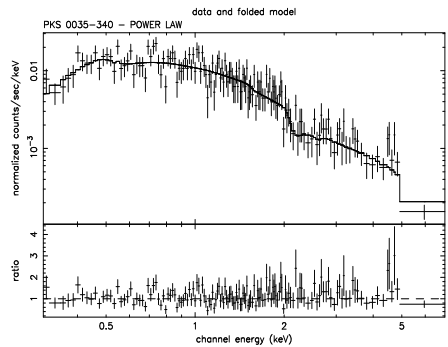}
\caption{PKS0035-340 spectrum fitted with a Power law model. The
excess at $\sim$ 4.5 keV is evident}
\label{pks}
\end{figure}

We investigate in detail the case of CXOJ003728.8-334442, the brightest
source detected in the area, which is 
positionally coincident with the radio source  \object{PKS 0035-340}, but so far
not identified. 
The statistics of the
X-ray data allows us to collect a spectrum, which we
bin in such a way that each channel has 
$\geq 2\sigma$ significance after background subtraction
(Fig. ~\ref{pks}).  A single power law model gives an acceptable
fit with a photon index $\Gamma$=1.72 [1.63-1.78] and 
$N_{H}$ = 6.1 [4.2-8.0] $\times 10^{20}$ cm$^{-2}$, consistent with
the classical AGN spectrum, and unabsorbed flux (0.5-10 keV band)
f$_X = 1.3 \times 10^{-13}$ erg cm$^{-2}$ s$^{-1}$.
As is evident from the figure, even if the $\chi^2$ is statistically
acceptable, there is an excess at  $\sim$ 4.5 keV.
If we interpret this feature as a Fe-K$\alpha$ line, we can use it to measure
the still unknown redshift of the source.
We can make two different hypotheses that imply two different
spectral models: that the source is an AGN (the radio
galaxy itself); or that the X-ray source is a cluster of galaxies, of
which the radio galaxy is a tracer (see e.g. Zanichelli et al. 2001, 
who however did not find a cluster around PKS0035-340).
In either case, the interpretation of the feature as the  Fe-K$\alpha$
line gives a similar z $\sim 0.45$.
An even better match with the spectral feature is obtained by fitting
both the 6.4 keV (``neutral'') and the 6.7 keV (``highly-ionized'') 
Fe-line at the same redshift. 
The significance of each line is only about 1$\sigma$; nevertheless
we obtain a consistent fit with z=0.425, and equivalent width of
$\sim 200$ eV for both lines ($< 500$ eV at the 90\% confidence level).
The derived luminosity would then be L$_X =8.5\times 10^{43}$ erg s$^{-1}$ 
(for H$_0$=75 km s$^{-1}$ Mpc$^{-1}$, or L$_X =1.9\times 10^{44}$ erg s$^{-1}$ 
for  H$_0$=50 km s$^{-1}$ Mpc$^{-1}$), consistent both with the AGN
and the cluster hypothesis.
The formal temperature fit of kT=6.8 [5.7-8.3] keV is broadly consistent
with the expectation from the luminosity -- temperature relation
of Markevitch (1998). 

To our knowledge this is the first redshift 
derived from X-ray spectroscopy without prior measures or estimates. 
The statistics is however
scanty and does not grant that the interpretation of the feature is correct.

We further analyzed the spectral data of  CXO J003747.4-334104 and  
CXO J003756.3-334124.

CXO J003747.4-334104  has a faint optical counterpart visible on the DSS II
red plate. 
A single power law model with Galactic N$_H$ gives an acceptable
fit with a photon index $\Gamma$=1.75 [1.66-1.85], consistent with
the canonical AGN spectrum, and flux (0.5-10 keV)
f$_X = 8.8 \times 10^{-14}$ erg cm$^{-2}$ s$^{-1}$. 
If the source is indeed an AGN with redshift in the range 0.4-2.0 (typical
of X-ray selected AGN) the derived luminosity would then be 
L$_X {\rm (0.5-10 keV)} = 0.5 - 20 \times 10^{44}$ erg s$^{-1}$ (for 
H$_0$=75 km s$^{-1}$ Mpc$^{-1}$), consistent with the AGN hypothesis.
If the small excess at $\sim 2.4$ keV were the Fe K$\alpha$ line, than the
redshift would be z $\sim 1.7$.

\begin{figure}
\psfig{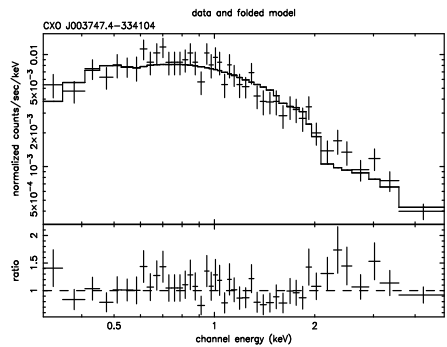}
\caption{The spectrum of CXO J003747.4-334104 fitted with a power law model. }
\label{cxo1}
\end{figure}

CXO J003756.3-334124 does not have an optical counterpart on the DSS II plate,
and therefore the most likely counterpart is a distant X-ray cluster.
The spectrum would then need to be fitted with a thermal or
plasma emission model; however, not knowing the redshift this could be tricky.
If we  use a bremsstrahlung model
the best fit temperature is kT =  9.4 [6.5-15.6] keV with a flux
(0.5-10 keV) f$_X = 1.3 \times 10^{-13}$ erg cm$^{-2}$ s$^{-1}$.    
A {\em mekal} model with abundance = 0.3 solar, at a few selected
redshifts between 0.5 and 1.2, also gives a reasonable fit 
and a similar flux.
For the same redshift range 0.5 to 1.2 (typical of X-ray selected clusters
of galaxies) the derived luminosity would then be 
L$_X (0.5-10 keV) = 1. - 8 \times 10^{44}$ erg s$^{-1}$ (for 
H$_0$=75 km s$^{-1}$ Mpc$^{-1}$), in the range of distant 
cluster luminosities.

A single power law could also fit the data within the errors,
and results in a slope $\Gamma = 1.42 [1.31 - 1.53]$, 
flatter than classical AGN spectra.  

\begin{figure}
\psfig{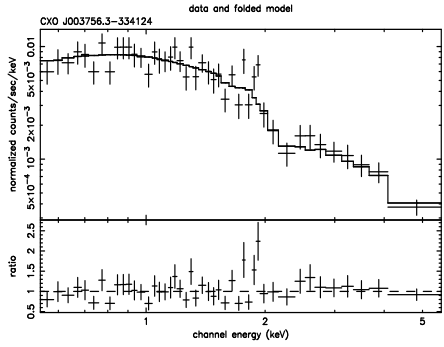}
\caption{The spectrum of CXO J003756.3-334124 fitted with a bremsstrahlung 
model. }
\label{cxo2}
\end{figure}

For these two sources no conclusion can be drawn from X-rays alone, and
a detailed optical investigation would be needed to define the counterpart.

\clearpage

\begin{table*}
{\scriptsize
\begin{center}
\tablefirsthead{
\hline
Name & RA  & Dec & Net cts & S/N  & Flux (erg cm$^{-2}$ s$^{-1}$) & Other names \\
CXOJ-&\multicolumn{2}{c}{J2000} & & &(0.5-10 keV)& \\
\hline
}
\tablehead{
\multicolumn{7}{l} {Table \ref{wavsrc} continued}\\
\hline
Name & RA  & Dec & Net cts & S/N  & Flux (erg cm$^{-2}$ s$^{-1}$) & Other names \\
CXOJ-&\multicolumn{2}{c}{J2000} & & &(0.5-10 keV)& \\
\hline
}
\tabletail{%
\hline \hline}
\tablelasttail{%
\hline \hline}
\tablecaption{Sources detected by the wavelet algorithm in the S3 field of view.
Count rates have been converted into flux by assuming a power law with 
index $\Gamma = 1.7 $ and Galactic low energy absorption N$_{H} =
2 \times 10^{20}$ cm$^{-2}$. An X-ray spectral analysis is presented in the
text for sources indicated with $^{*}$.}
\label{wavsrc}
\begin{supertabular}{| l r r r r r r|}
\hline

 003724.1-334309 & 0 37 24.14 &-33 43 09.02&  131.24 $\pm$  11.62&  42.03& 1.3$\times 10^{-14}$& \\
 003727.1-334245 & 0 37 27.11 &-33 42 45.61&    6.17 $\pm$   2.65&   2.73& 5.9$\times 10^{-16}$& \\
 003727.6-334347 & 0 37 27.61 &-33 43 47.78&   10.97 $\pm$   3.46&   4.70& 1.0$\times 10^{-15}$& \\
 003728.2-334407 & 0 37 28.26 &-33 44 07.18&   49.61 $\pm$   7.21&  17.89& 4.7$\times 10^{-15}$& \\
 003728.8-334442$^{*}$ & 0 37 28.83 &-33 44 42.26& 1500.78 $\pm$  38.92& 308.23& 1.4$\times 10^{-13}$& PKS 0035-340  \\
 003729.6-334638 & 0 37 29.68 &-33 46 38.82&    6.73 $\pm$   3.00&   2.46& 6.4$\times 10^{-16}$& \\
 003731.5-334404 & 0 37 31.55 &-33 44 04.83&   15.42 $\pm$   4.12&   6.11& 1.5$\times 10^{-15}$& \\
 003733.0-334338 & 0 37 33.00 &-33 43 38.41&   21.40 $\pm$   4.80&   8.45& 2.0$\times 10^{-15}$& \\
 003733.6-334543 & 0 37 33.69 &-33 45 43.68&   12.63 $\pm$   3.74&   5.14& 1.2$\times 10^{-15}$& \\
 003734.0-334601 & 0 37 34.06 &-33 46 01.24&   12.66 $\pm$   3.74&   5.18& 1.2$\times 10^{-15}$& \\
 003735.1-334517 & 0 37 35.14 &-33 45 17.32&  247.79 $\pm$  15.94&  68.12& 2.4$\times 10^{-14}$& \\
 003737.3-334515 & 0 37 37.33 &-33 45 15.88&   27.06 $\pm$   5.39&  10.25& 2.6$\times 10^{-15}$& \\
 003737.5-334342 & 0 37 37.54 &-33 43 42.09&    8.12 $\pm$   3.00&   3.56& 7.8$\times 10^{-16}$& \\
 003737.5-334256 & 0 37 37.58 &-33 42 56.94&  116.10 $\pm$  10.95&  36.78& 1.1$\times 10^{-14}$& N.17 \\
 003737.6-334254 & 0 37 37.60 &-33 42 54.95&   94.51 $\pm$   9.90&  30.90& 9.0$\times 10^{-15}$& N.16 \\
 003738.3-334308 & 0 37 38.32 &-33 43 08.72&   16.06 $\pm$   4.24&   6.08& 1.5$\times 10^{-15}$& N.15 \\
 003738.7-334316 & 0 37 38.71 &-33 43 16.08&  105.92 $\pm$  10.54&  31.01& 1.0$\times 10^{-14}$& N.14 \\
 003738.7-334318 & 0 37 38.78 &-33 43 18.66&   38.02 $\pm$   6.40&  12.96& 3.6$\times 10^{-15}$& N.13 \\
 003738.9-334517 & 0 37 38.98 &-33 45 17.73&   14.26 $\pm$   4.00&   5.53& 1.4$\times 10^{-15}$& \\
 003739.0-334118 & 0 37 39.06 &-33 41 18.93&   22.50 $\pm$   4.90&   9.00& 2.1$\times 10^{-15}$& \\
 003739.1-334229 & 0 37 39.15 &-33 42 29.57&   43.75 $\pm$   6.78&  16.02& 4.2$\times 10^{-15}$& N.12 \\
 003739.1-334123 & 0 37 39.16 &-33 41 23.42&    7.74 $\pm$   3.00&   3.20& 7.4$\times 10^{-16}$& \\
 003739.2-334250 & 0 37 39.21 &-33 42 50.10&   62.59 $\pm$   8.06&  22.54& 6.0$\times 10^{-15}$& N.11 \\
 003739.3-334323 & 0 37 39.38 &-33 43 23.07&  383.77 $\pm$  19.77& 100.35& 3.7$\times 10^{-14}$& N.10 \\
 003740.1-334326 & 0 37 40.15 &-33 43 26.01&    5.16 $\pm$   2.45&   2.28& 4.9$\times 10^{-16}$& N.24 \\
 003740.4-334324 & 0 37 40.44 &-33 43 24.87&   14.28 $\pm$   4.00&   5.56& 1.4$\times 10^{-15}$& N.22, NVSS J003740-334324  \\
 003740.4-334013 & 0 37 40.45 &-33 40 13.11&   71.40 $\pm$   8.60&  25.24& 6.8$\times 10^{-15}$& \\
 003740.8-334330 & 0 37 40.86 &-33 43 30.85&   66.20 $\pm$   8.31&  22.96& 6.3$\times 10^{-15}$& N.9 \\
 003741.0-334221 & 0 37 41.03 &-33 42 21.37&   21.23 $\pm$   4.80&   8.20& 2.0$\times 10^{-15}$& N.8 \\
 003741.0-334331 & 0 37 41.08 &-33 43 31.81&   70.64 $\pm$   8.60&  23.34& 6.7$\times 10^{-15}$& N.7 \\
 003741.2-334232 & 0 37 41.23 &-33 42 32.08&    6.97 $\pm$   2.83&   2.99& 6.7$\times 10^{-16}$& N.21 \\
 003741.9-334326 & 0 37 41.98 &-33 43 26.36&    8.54 $\pm$   3.16&   3.43& 8.2$\times 10^{-16}$& N.23 \\
 003742.1-334313 & 0 37 42.11 &-33 43 13.72&   13.20 $\pm$   3.87&   5.08& 1.3$\times 10^{-15}$& N.20 \\
 003742.3-334122 & 0 37 42.39 &-33 41 22.80&    7.92 $\pm$   3.00&   3.37& 7.6$\times 10^{-16}$& \\
 003742.4-334304 & 0 37 42.47 &-33 43 04.18&   19.97 $\pm$   4.69&   7.48& 1.9$\times 10^{-15}$& N.6 \\
 003742.7-334212 & 0 37 42.79 &-33 42 12.46&   40.58 $\pm$   6.63&  13.34& 3.9$\times 10^{-15}$& N.5, G1\\
 003742.8-334209 & 0 37 42.84 &-33 42 09.63&   18.57 $\pm$   4.58&   6.67& 1.8$\times 10^{-15}$& N.19, G1\\
 003743.0-334020 & 0 37 43.05 &-33 40 20.75&   21.09 $\pm$   4.80&   8.02& 2.0$\times 10^{-15}$& \\
 003743.0-334205 & 0 37 43.01 &-33 42 05.96&   48.56 $\pm$   7.14&  17.43& 4.6$\times 10^{-15}$& N.4, G1\\
 003743.1-334142 & 0 37 43.10 &-33 41 42.91&   12.45 $\pm$   3.74&   4.94& 1.2$\times 10^{-15}$& \\
 003743.1-334203 & 0 37 43.12 &-33 42 03.96&   72.32 $\pm$   8.72&  23.30& 6.9$\times 10^{-15}$& N.3, G1\\
 003743.3-334912 & 0 37 43.32 &-33 49 12.53&  224.16 $\pm$  16.79&  25.90& 2.1$\times 10^{-14}$& \\
 003743.3-334312 & 0 37 43.35 &-33 43 12.62&    6.82 $\pm$   2.83&   2.86& 6.5$\times 10^{-16}$& N.18 \\
 003743.6-334147 & 0 37 43.69 &-33 41 47.24&   14.43 $\pm$   4.00&   5.71& 1.4$\times 10^{-15}$& \\
 003743.7-334534 & 0 37 43.76 &-33 45 34.41&   22.94 $\pm$   5.00&   8.58& 2.2$\times 10^{-15}$& \\
 003743.8-334209 & 0 37 43.85 &-33 42 09.63&   39.87 $\pm$   6.56&  13.42& 3.8$\times 10^{-15}$& N.2, G1\\
 003744.0-334028 & 0 37 44.00 &-33 40 28.72&   52.58 $\pm$   7.42&  18.90& 5.0$\times 10^{-15}$& \\
 003745.2-334228 & 0 37 45.29 &-33 42 28.31&   73.21 $\pm$   8.77&  23.38& 7.0$\times 10^{-15}$& N.1, near G2\\
 003745.6-334151 & 0 37 45.61 &-33 41 51.85&  233.81 $\pm$  15.39&  78.32& 2.2$\times 10^{-14}$& \\
 003745.7-334546 & 0 37 45.77 &-33 45 46.64&   31.46 $\pm$   5.83&  11.18& 3.0$\times 10^{-15}$& \\
 003747.0-333952 & 0 37 47.07 &-33 39 52.20&    9.61 $\pm$   3.32&   3.90& 9.2$\times 10^{-16}$& G3\\
 003747.4-334104$^{*}$ & 0 37 47.44 &-33 41 04.16&  958.12 $\pm$  31.08& 243.33& 9.2$\times 10^{-14}$& \\
 003748.6-333841 & 0 37 48.67 &-33 38 41.63&    3.09 $\pm$   2.00&   1.35& 3.0$\times 10^{-16}$& \\
 003748.8-334438 & 0 37 48.80 &-33 44 38.51&  130.79 $\pm$  11.62&  40.51& 1.2$\times 10^{-14}$& \\
 003749.2-334404 & 0 37 49.20 &-33 44 04.71&  367.52 $\pm$  19.31& 105.15& 3.5$\times 10^{-14}$& \\
 003749.3-334107 & 0 37 49.36 &-33 41 07.58&   85.18 $\pm$   9.38&  29.49& 8.1$\times 10^{-15}$& \\
 003750.0-334641 & 0 37 50.01 &-33 46 41.14&  414.66 $\pm$  20.88&  72.74& 4.0$\times 10^{-14}$& \\
 003751.2-334659 & 0 37 51.23 &-33 46 59.26&   14.39 $\pm$   4.24&   4.66& 1.4$\times 10^{-15}$& \\
 003751.2-334216 & 0 37 51.25 &-33 42 16.52&   12.45 $\pm$   3.74&   4.95& 1.2$\times 10^{-15}$& \\
 003751.4-334135 & 0 37 51.41 &-33 41 35.05&   28.84 $\pm$   5.57&  10.65& 2.8$\times 10^{-15}$& \\
 003751.6-334100 & 0 37 51.65 &-33 41 00.24&  109.36 $\pm$  10.68&  32.91& 1.0$\times 10^{-14}$& \\
 003752.4-334403 & 0 37 52.44 &-33 44 03.45&  155.06 $\pm$  12.65&  45.80& 1.5$\times 10^{-14}$& \\
 003753.1-334131 & 0 37 53.16 &-33 41 31.97&   14.37 $\pm$   4.00&   5.65& 1.4$\times 10^{-15}$& \\
 003754.1-334630 & 0 37 54.16 &-33 46 30.93&   64.90 $\pm$   8.66&  15.12& 6.2$\times 10^{-15}$& \\
 003754.4-334630 & 0 37 54.43 &-33 46 30.75&   56.47 $\pm$   7.94&  15.27& 5.4$\times 10^{-15}$& \\
 003754.5-334442 & 0 37 54.54 &-33 44 42.30&   13.65 $\pm$   4.36&   3.93& 1.3$\times 10^{-15}$& \\
 003755.2-334159 & 0 37 55.28 &-33 41 59.45&   16.14 $\pm$   4.24&   6.17& 1.5$\times 10^{-15}$& \\
 003755.7-334412 & 0 37 55.71 &-33 44 12.00&   21.81 $\pm$   5.20&   6.34& 2.1$\times 10^{-15}$& \\
 003755.7-334755 & 0 37 55.75 &-33 47 55.05&    8.09 $\pm$   4.00&   2.05& 7.7$\times 10^{-16}$& \\
 003756.3-334224$^{*}$ & 0 37 56.32 &-33 42 24.76&  992.62 $\pm$  31.72& 208.57& 9.5$\times 10^{-14}$& \\
 003756.0-334221 & 0 37 56.05 &-33 42 21.63&   17.04 $\pm$   4.36&   6.44& 1.6$\times 10^{-15}$& \\
 003801.2-334430 & 0 38 01.26 &-33 44 30.16&   26.32 $\pm$   5.74&   7.06& 2.5$\times 10^{-15}$& \\
\hline
\end{supertabular}
\end{center}
}
\end{table*}

\listofobjects
\end{document}